\documentclass[11pt]{article}

\usepackage[utf8]{inputenc}
\usepackage[english]{babel}

\usepackage{amsmath, bm}
\usepackage{amssymb}
\usepackage{amsthm}
\usepackage{graphicx}
\usepackage[shortlabels]{enumitem}
\usepackage{accents, setspace}
\usepackage{xcolor}
\usepackage{hyperref}
\usepackage{titlesec}
\usepackage{listings}
\usepackage[round]{natbib}
\usepackage[papersize={8.5in, 11in}, margin=1.2in]{geometry}

\newtheorem{theorem}{Theorem}

\newtheorem*{theorem*}{Theorem}

\newcommand{\keyword}{\textit{Keywords: }}
\newcommand{\utilde}{\underaccent{\tilde}}
\newcommand{\btheta}{\boldsymbol{\theta}}
\newcommand{\bthetan}{\widehat{\boldsymbol{\theta}}_n}

\theoremstyle{definition}

\usepackage{graphicx}
\graphicspath{ {./figures/} }

\titlespacing{\subsubsection}{0pt}{\parskip}{-\parskip}

\usepackage{fancyhdr}
\begin{document}

\title{Methods to Compute Prediction Intervals:\\A Review and New Results}
\author{Qinglong Tian, Daniel J. Nordman, William Q. Meeker\\
\\
\normalsize{Department of Statistics, Iowa State University, Ames, IA 50011}
}
\maketitle

\begin{abstract}

The purpose of this paper is to review both classic and modern methods for constructing prediction intervals.
We focus, primarily, on model-based non-Bayesian methods for the prediction of a scalar random variable, but we also include Bayesian methods with objective prior distributions.
Our review of non-Bayesian methods follows two lines: general methods based on (approximate) pivotal quantities and methods based on non-Bayesian predictive distributions.
The connection between these two types of methods is described for distributions in the (log-)location-scale family.
We also discuss extending the general prediction methods to data with complicated dependence structures as well as some non-parametric prediction methods (e.g., conformal prediction).

\end{abstract}

\noindent\keyword{bootstrap; calibration; coverage probability; prediction interval; predictive distribution}

\section{Introduction}

\subsection{Prediction History and Notation}\label{subsec:history}

While most statistics textbooks and courses emphasize the explanatory and descriptive roles of statistics, the topic of statistical prediction often receives less attention, despite its practical importance, as noted in recent commentaries (cf.~\citealt{shmueli2010explain} and \citealt{harville2014need}).
The goal of this paper is to review important prediction interval methods and show some interesting connections.
We start with some history of statistical prediction.

\subsubsection{Bayesian Prediction}

Bayesian prediction is accomplished via the Bayesian predictive distribution, which is a conditional distribution of future random variables given the observed data.
When the future random variable $Y$ is conditionally independent of the sample $\boldsymbol{X}_n$ given $\btheta$, the Bayesian predictive distribution, in the form of probability density function (pdf), is computed as
\begin{equation}\label{eq:bayesian-predictive-distribution}
p(y|\boldsymbol{x}_n)=\int p(y|\btheta)p(\btheta|\boldsymbol{x}_n)d\btheta,
\end{equation}
where $p(y|\btheta)$ is the pdf of $Y$ conditional on $\btheta$ and $p(\btheta|\boldsymbol{x}_n)$ is the posterior distribution of $\btheta$ given the realized value $\boldsymbol{x}_n$ of the random data $\boldsymbol{X}_n$.
Similar to a credible interval, a $100(1-\alpha)\%$ Bayesian prediction interval can be obtained from the $\alpha/2$ and $1-\alpha/2$ quantiles of the Bayesian predictive distribution.

Bayesian prediction can be traced back to Laplace's 1774 Memoir which contains a derivation of the Bayesian predictive distribution for a binomial random variable (cf.~\citealt{stigler1986laplace}).
The early work of de Finetti (e.g., \citealt{de1937prevision}) is a cornerstone of Bayesian statistics wherein the importance of the Bayesian predictive distribution under de Finetti's subjective viewpoint of probability is emphasized.
\citet{gkisser2017predictive} states that ``The completely observabilistic view was brought particularly to the attention of British and American statisticians with the translation into English of the book of probability authored by \citet{de2017theory},'' where the ``completely observabilistic view'' refers to the principle of assigning probability only to observable events.
\citet{fortini2014predictive} provide a review of the Bayesian predictive distribution, noting that ``In the \textit{predictive approach} (also referred to as the \textit{generative model}), the uncertainty is directly described through the predictive distributions.''

During the 1960s and 1970s, more work was done to apply Bayesian prediction methods.
For example, \cite{guttman1964bayesian} use a Bayesian predictive distribution to solve ``best population problems,'' while \citet{thatcher1964relationships} revisits the binomial prediction problem considered by Laplace and compares Bayesian and non-Bayesian prediction methods.
\citet{aitchison_dunsmore_1975} and \citet{gkisser2017predictive} are books that describe methods for Bayesian prediction.

Although Bayesian statistical methods are not constrained to specific models and distributions, they would wait for the rediscovery of MCMC sampling methods in the late 1980s to take advantage of that generality beyond special distributions that have conjugate prior distributions, such as distributions in the natural exponential family (e.g., the normal, binomial, and Poisson distributions).
Also, the large increase in computing power over past decades and computer software like Stan have facilitated the application of Bayesian methods to more complicated prediction problems.

\subsubsection{Non-Bayesian Prediction}\label{subsub:non-bayesian}

Similar to the confidence interval for parameters and functions of parameters, prediction intervals provide a natural approach for quantifying prediction uncertainty.
Let $\boldsymbol{X}_n$ be the sample and $Y$ be the future random variable.
An \textit{exact} $100(1-\alpha)\%$ prediction interval method for $Y$, denoted by $\text{PI}(\boldsymbol{X}_n)$, satisfies
$
\Pr{}_{\!\btheta}[Y\in\text{PI}(\boldsymbol{X}_n)]=1-\alpha
$,
where $\btheta$ contains the parameters that index the joint distribution of $(\boldsymbol{X}_n,Y)$, and $1-\alpha$ is called the nominal confidence level because not all prediction interval methods are {\em exact}.
If $\Pr{}_{\!\btheta}[Y\in\text{PI}(\boldsymbol{X}_n)]\geq1-\alpha$ the procedure is said to be {\em conservative}.
If $\Pr{}_{\!\btheta}[Y\in\text{PI}(\boldsymbol{X}_n)]\to1-\alpha$ as $n\to\infty$, we say the method is \textit{asymptotically correct}.
It is worth noting that the terminology is different in some literature (especially in nonparametric literature), where the term ``exact'' is used to indicate being conservative.

In one of the earliest works on non-Bayesian prediction, \citet{fisher1935fiducial} uses a fiducial method to construct a prediction interval for a new observation and a future sample mean from a normal distribution, given a previous sample from the same distribution.
Around the same time, \citet{baker1935probability} considers predicting a future sample mean and implicitly provides the correct frequentist interpretation for Fisher's interval.
In a paper describing sampling sizes for setting tolerance limits, \citet{Wilks1941} also gives Fisher's formula and refers to it as limits that will ``\dots include on the average a proportion $a$ of the universe between them\dots,'' which would later be called a $\beta$-expectation tolerance interval (equivalent to a prediction interval).

The first use of the term ``prediction interval'' seems to have come somewhat later.
Using a frequentist approach, \citet{proschan1953confidence} derives the same interval as Fisher and writes ``such an interval might be called more appropriately a prediction interval, since the term `confidence interval' generally refers to population parameters.''
\citet{thatcher1964relationships} investigates binomial distribution prediction but used ``confidence limit for the prediction'' to refer to the prediction interval.
As documented in \citet{patel1989prediction}, starting in the late 1960s, numerous papers began appearing in engineering and applied statistics journals presenting methods for many specific prediction problems and using the term ``prediction interval.''

In the following decades, statisticians began to develop general methods to construct prediction intervals.
These include methods based on pivotal quantities, fiducial distributions, and non-Bayesian predictive distributions.
Recent developments with these approaches often involve resampling and other simulation-based approximations.
These general methods will be described and illustrated in the rest of this paper.

\subsection{Overview}

In this paper, we focus on constructing prediction intervals for problems where the predictand $Y$ is a scalar and a parametric model (with unknown parameters) is used to describe the distributions of $Y$ and the data $\boldsymbol{X}_n$.
We mainly consider the cases where $\boldsymbol{X}_n$ and $Y$ are generated through a random sampling process or a slight variant of it.
We describe general non-Bayesian methods for prediction that have been proposed in this setting.
We view prediction and prediction interval methods from a frequentist perspective where methods are evaluated, primarily, on the basis of coverage probability, relative to a specified nominal confidence level.
Although we assess prediction methods using frequentist criteria, our review includes Bayesian methods with non-informative (or objective) prior distributions.
In fact, Bayesian methods with non-informative prior distributions provide an important means of defining prediction methods for both complicated and simple statistical models.
Such Bayesian-based methods have been shown to have good frequentist properties (i.e., coverage probability close to nominal; e.g., \citealt{hulting1991some}, \citealt{harville1992classical}, and Section~\ref{sec:location-scale} of this paper).

The remainder of this paper is organized as follows.
Section~\ref{sec:pred-interval} describes methods to construct a prediction interval based on pivot-type relations.
Section~\ref{sec:pred-dist-concept} discusses the concept of a (non-Bayesian) predictive distribution as an alternative but equivalent approach to prediction intervals, and Section~\ref{sec:pred-dist-method} outlines several methods to construct predictive distributions.
Section~\ref{sec:location-scale} applies some prediction methods to the (log-)location-scale distribution family and provides new results on connections among various prediction methods.
Section~\ref{sec:examples} describes how to apply general prediction methods to other continuous distributions and provides two illustrative examples.
Section~\ref{sec:discrete} describes several general methods that can be applied to construct prediction intervals for discrete distributions.
While Sections~\ref{sec:pred-interval}--\ref{sec:discrete} primarily focus on independent data or data with a simple dependence structure, Section~\ref{sec:dependent-data} discusses extensions involving prediction methods when $Y$ and $\boldsymbol{X}_n$ have a complicated dependence structure.
Section~\ref{sec:model-free} discusses several model-free prediction methods.
Section~\ref{sec:conclusion} provides some concluding remarks.

\section{Prediction Interval Methods}\label{sec:pred-interval}

\subsection{Pivotal Methods}\label{subsec:pivotal-method}

\citet{cox1975} describes the pivotal prediction method, where the main idea is to find a scalar quantity $q(\boldsymbol{X}_n,Y)$ that does not depend on any parameters.
Then the $1-\alpha$ prediction set of $Y$ is given by
\begin{equation}\label{eq:pivotal-method}
\left\{y:q(\boldsymbol{x}_n,y)\leq q_{n,1-\alpha}\right\},
\end{equation}
where $q_{n,1-\alpha}$ is the $1-\alpha$ quantile of $q(\boldsymbol{X}_n,Y)$.
If $q(\boldsymbol{x}_n,y)$ is a monotone function of $y$, the prediction region (\ref{eq:pivotal-method}) becomes a one-sided prediction bound.
When $q(\boldsymbol{X}_n,Y)$ is continuous, the pivotal prediction method is exact because $\Pr_{\btheta}\left[q(\boldsymbol{X}_n,Y)\leq q_{n,1-\alpha}\right]=1-\alpha$ for any $\alpha\in(0,1)$.
The rest of this section describes two special types of pivotal methods.
\subsubsection{Inverting a Hypothesis Test}
\citet{cox1975} suggests a prediction method based on inverting a hypothesis test.
Suppose $\boldsymbol{X}_n\sim f(x;\btheta)$ and that $Y\sim f(y;\btheta^\dagger)$ is independent of $\boldsymbol{X}_n$.
Let $w_{\alpha}$ be a size $\alpha$ critical region for a similarity test $\btheta=\btheta^\dagger$; that is, $\Pr{}_{\!\boldsymbol{\theta}=\boldsymbol{\theta}^\dagger}\left[(\boldsymbol{X}_n, Y)\in w_\alpha\right]=\alpha$, where $\Pr{}_{\!\boldsymbol{\theta}=\boldsymbol{\theta}^\dagger}(\cdot)$ denotes any probability function that belongs to the subset $\{\Pr_{\!(\btheta,\btheta^\dagger)}:\btheta=\btheta^\dagger\}$.
Then a $1-\alpha$ prediction region is defined as $\{y:(\boldsymbol{x}_n,y)\not\in w_{\alpha}\}$.
\subsubsection{Using the Probability Integral Transform}
\label{subsubsec:pivotal-conditional-cdf-method}

If $T(\boldsymbol{X}_n)$ is a scalar statistic with a continuous cumulative distribution function (cdf) $F(\cdot;\btheta)$, then $F[T(\boldsymbol{X}_n);\theta]$ has a $\text{Uniform}(0,1)$ distribution where $F(\cdot;\theta)$ is the cdf of $T(\boldsymbol{X}_n)$.
The pivotal cdf method uses the probability integral transform to compute confidence intervals for a scalar parameter (e.g., \citealt[Chapter 9]{casella2002statistical}).
When $F(\cdot;\theta)$ is a monotone function of $\theta$, a $1-\alpha$ equal-sided confidence interval is given by $\left\{\theta:\alpha/2\leq F[T(\boldsymbol{X}_n;\theta)]\leq1-\alpha/2\right\}$.

The pivotal cdf method can be extended to define prediction interval methods.
Let $T(\boldsymbol{X}_n)$ be a statistic from data $\boldsymbol{X}_n$ and $R(\boldsymbol{X}_n,Y)$ be a statistic from both the data and the predictand.
When the conditional cdf of $T(\boldsymbol{X}_n)$ given $R(\boldsymbol{X}_n, Y)$, say $G_{T|R}\left[t|R(\boldsymbol{x}_n, y)\right]$, does not depend on any parameters and is continuous, then $G_{T|R}\left[T(\boldsymbol{X}_n)|R(\boldsymbol{X}_n, Y)\right]$ has a $\text{Uniform}(0, 1)$ distribution.
If $G_{T|R}\left[t|R(\boldsymbol{x}_n, y)\right]$ is a non-increasing function of $y$, then $1-\alpha$ lower and upper prediction bounds are defined as
\begin{equation}\label{eq:conditional-cdf-bounds}
	\undertilde{Y}_{1-\alpha}=\inf\{y:G_{T|R}\left[T(\boldsymbol{x}_n)|R(\boldsymbol{x}_n, y)\right]<1-\alpha\},\quad
	\widetilde{Y}_{1-\alpha}=\sup\{y:G_{T|R}\left[T(\boldsymbol{x}_n)|R(\boldsymbol{x}_n, y)\right]>\alpha\}.
\end{equation}
Because $G_{T|R}\left[T(\boldsymbol{X}_n)|R(\boldsymbol{X}_n, Y)\right]\sim\text{Uniform}(0,1)$, we have $\Pr(Y\geq\undertilde{Y}_{1-\alpha})=\Pr(Y\leq\widetilde{Y}_{1-\alpha})=1-\alpha$.
When $G_{T|R}\left[t|R(\boldsymbol{x}_n;y)\right]$ is non-decreasing in $y$, then
\[
\undertilde{Y}_{1-\alpha}=\inf\{y:G_{T|R}\left[T(\boldsymbol{x}_n)|R(\boldsymbol{x}_n, y)\right]>\alpha\},\quad
\widetilde{Y}_{1-\alpha}=\sup\{y:G_{T|R}\left[T(\boldsymbol{x}_n)|R(\boldsymbol{x}_n, y)\right]<1-\alpha\}.
\]
Similar, but conservative, prediction methods can also be formulated with discrete cdfs.
Section~\ref{sub-sec-piv-cdf-methods} describes such applications to discrete distributions with a scalar parameter.

\subsection{Approximate Pivotal Methods}
\label{subsec:approximate-pivotal-methods}
Suppose $q(\boldsymbol{X}_n,Y)$ (with cdf $Q_n(\cdot;\btheta)$) is a quantity that converges in distribution to a pivotal quantity (with cdf $Q(\cdot)$).
In the absence of a pivotal quantity, the approximate pivotal quantity $q(\boldsymbol{X}_n, Y)$ can also be used to construct a prediction interval.

Because $G(Y|\boldsymbol{X}_n;\btheta)$ is $\text{Uniform}(0,1)$ distributed when $Y$ given $\boldsymbol{X}_n$ is continuous with conditional cdf $G(\cdot|\boldsymbol{X}_n;\btheta)$, an approximate pivotal quantity that is available in most cases is $U_n\equiv G(Y|\boldsymbol{X}_n;\bthetan)$, which converges in distribution to $\text{Uniform}(0, 1)$ if $\bthetan$ is a consistent estimator of $\btheta$, usually the maximum likelihood (ML) estimator.
Letting $u_{n,1-\alpha}$ be the $1-\alpha$ quantile of $U_n$, we have $1-\alpha=\Pr{}_{\!\btheta}(U_n\leq u_{n,1-\alpha})=\Pr{}_{\!\btheta}[Y\leq G^{-1}(u_{n,1-\alpha}|\boldsymbol{X}_n;\widehat{\boldsymbol{\theta}}_n)]$, where $G^{-1}(\cdot|\boldsymbol{X}_n;\btheta)$ is the quantile function of $Y$ given $\boldsymbol{X}_n$.
Because $u_{n,1-\alpha}$ often depends on the unknown parameter $\btheta$, an estimate of $u_{n,1-\alpha}$ can be used instead.
The rest of this section describes three ways to estimate $u_{n,1-\alpha}$.

\subsubsection{The Plug-in Method}\label{subsubsec:plug-in}

The plug-in method, also known as the naive or estimative method, is to use $1-\alpha$ (i.e., the $1-\alpha$ quantile of the $\text{Uniform}(0,1)$ distribution) to replace $u_{n,1-\alpha}$.
The plug-in $1-\alpha$ upper prediction bound is defined as $\{y:y\leq y_{1-\alpha}(\bthetan,\boldsymbol{X}_n)\}$, where $y_{1-\alpha}(\bthetan,\boldsymbol{X}_n)\equiv\inf\{y: G(y|\boldsymbol{X}_n;\bthetan)\geq1-\alpha\}$.
The $1-\alpha$ lower bound can be defined as $\{y:y\geq y_\alpha(\bthetan,\boldsymbol{X}_n)\}$.
Operationally, the plug-in method replaces the unknown parameters $\btheta$ with a consistent estimator $\bthetan$ in the quantile $y_{1-\alpha}(\btheta,\boldsymbol{X}_n)\equiv\inf\{y: G(y|\boldsymbol{X}_n;\btheta)\geq1-\alpha\}$.
The coverage probability of the plug-in method is typically different from the nominal confidence level because the sampling error in $\bthetan$ is ignored.
Under certain regularity conditions, the error of the coverage probability of the plug-in method is of order $O(1/n)$ (cf.~\citealt{cox1975}, \citealt{beran1990}, \citealt{hall1999}).

\subsubsection{Calibration-Bootstrap Method}\label{subsubsec:calibration-bootstrap}

To reduce the plug-in coverage error, \citet{beran1990} proposes the calibration-bootstrap method.
Instead of using $1-\alpha$ to estimate $u_{n,1-\alpha}$, a bootstrap re-creation of this quantile is used.
The cdf of $U_n=G(Y|\boldsymbol{X}_n;\bthetan)$ is denoted by $H_n(\cdot;\btheta)$, and $H_n^{-1}(1-\alpha;\bthetan)$ is used to estimate $u_{n,1-\alpha}$, where $H^{-1}_n(\cdot;\btheta)$ is the quantile function of $U_n$.
Then, the $1-\alpha$ upper prediction bound using the calibration-bootstrap method is given as $y_{H_n^{-1}(1-\alpha;\bthetan)}(\bthetan)=\inf\{y:G(y;\bthetan)\geq H_n^{-1}(1-\alpha;\bthetan)\}$.
When a closed-form expression for $H_n(\cdot;\cdot)$ is not available, the bootstrap method can be used to approximate $H_n^{-1}(1-\alpha;\bthetan)$.
The bootstrap procedure is as follows,
\begin{enumerate}[topsep=0pt,itemsep=-1ex,partopsep=1ex,parsep=1ex]
	\item Generate a bootstrap sample $\boldsymbol{x}_n^\ast$ from the cdf $F(\cdot; \widehat{\boldsymbol{\theta}}_n)$.
	\item Compute a bootstrap estimate $\widehat{\boldsymbol{\theta}}_n^\ast$ of $\boldsymbol{\theta}$ using the bootstrap sample $\boldsymbol{x}_n^\ast$.
	\item Generate $y^\ast$, which is the bootstrap version of $Y$, from the cdf $G(\cdot|\boldsymbol{x}_n^\ast; \widehat{\boldsymbol{\theta}}_n)$.
	\item Compute $u^\ast = G(y^\ast|\boldsymbol{x}_n^\ast;\widehat{\boldsymbol{\theta}}_n^\ast)$.
	\item Repeat the above steps $B$ times to obtain a collection $\{u^\ast_1,\dots,u^\ast_B\}$ and define $\widetilde{u}_{1-\alpha}$ as the $1-\alpha$ sample quantile of these values.
	\item The $1-\alpha$ upper calibration prediction bound is $G^{-1}(\widetilde{u}_{1-\alpha}|\boldsymbol{x}_n;\widehat{\boldsymbol{\theta}}_n)$.
\end{enumerate}
\citet{beran1990} proves that, under regularity conditions, the error of coverage probability of the calibration-bootstrap method is of order $O(1/n^2)$, which is faster than the plug-in method rate $O(1/n)$.

\subsubsection{Calibration Using an Asymptotic Expansion}
\label{subsec:calibration-expansion}
Another method to improve on the plug-in method is to use asymptotic expansion (cf.,~\citealt{cox1975}, \citealt{barncox1996}, \citealt{vidoni1998note}).
To simplify the presentation, we illustrate this method under the assumption that $\boldsymbol{X}_n$ are independent of $Y$.
Let $\bthetan$ be an estimator of $\btheta$ that satisfies
\[
\text{E}_{\btheta}(\bthetan) = \boldsymbol{\theta}+a(\boldsymbol{\theta})/n+
o\left(\frac{1}{n}\right),\qquad\text{Var}_{\btheta}(\bthetan) = b(\boldsymbol{\theta})
/n+o\left(\frac{1}{n}\right).
\]
Because $\boldsymbol{X}_n$ and $Y$ are independent, we denote the $1-\alpha$ quantile of $Y$ by $y_{1-\alpha}(\btheta)=\inf\{y:G(y;\btheta)\geq 1-\alpha\}$, where $G(\cdot;\btheta)$ is the cdf of $Y$; we further define $\kappa_{\alpha,\btheta}(\bthetan)\equiv\Pr{}_{\!\btheta}\{Y\leq y_{1-\alpha}[\bthetan(\boldsymbol{X}_n)]\}$.
Under smoothness conditions, $\kappa_{\alpha,\btheta}(\bthetan)$ can be approximated using a Taylor-series expansion around $\btheta$ so that, upon taking expectations, the coverage probability of the plug-in prediction bound can be expressed as
\begin{equation}
	\begin{split}
		\Pr{}_{\!\btheta}\left[Y\leq y_{1-\alpha}(\widehat{\boldsymbol{\theta}}_n)\right]=&
		\text{E}_{\btheta}\left\{\kappa_{\alpha,\btheta}(\widehat{\boldsymbol{\theta}}_n)\right\}\\
		=&\text{E}_{\btheta}\left\{\kappa_{\alpha,\btheta}(\boldsymbol{\theta})+(\widehat{
			\boldsymbol{\theta}}_n-\boldsymbol{\theta})\kappa_{\alpha,\btheta}^\prime(
		\boldsymbol{\theta})+\frac{1}{2}(\widehat{\boldsymbol{\theta}}_n-
		\boldsymbol{\theta})^2\kappa_{\alpha,\btheta}^{\prime\prime}(\boldsymbol{\theta})
		\right\}+o\left(\frac{1}{n}\right)\\
		=&1-\alpha+c(\boldsymbol{\theta})/n+
		o\left(\frac{1}{n}\right),
	\end{split}
	\label{eq:calibration-expansion}
\end{equation}
for $c(\boldsymbol{\theta})\equiv a(\boldsymbol{\theta})\kappa_{\alpha,\btheta}^
{\prime}(\boldsymbol{\theta})+b(\boldsymbol{\theta})\kappa_{\alpha,\btheta}^{\prime\prime}(\boldsymbol{\theta})/2$ depending on the bias and variance of $\bthetan$ as well as the first derivative $\kappa^\prime_{\alpha,\btheta}(\btheta)$ and second derivatives $\kappa^{\prime\prime}_{\alpha,\btheta}(\btheta)$ of $\kappa_{\alpha,\btheta}(\btheta)$.
Letting $\alpha_c=\alpha+c(\btheta)/n$, then by replacing $1-\alpha$ with $1-\alpha_c$ in (\ref{eq:calibration-expansion}) we have
\begin{equation*}
	\begin{split}
		\Pr{}_{\!\btheta}\left[Y\leq y_{1-\alpha_c}(\bthetan)\right]=&\text{E}{}_{\btheta}\left\{1-\alpha_c+(\bthetan-\boldsymbol{\theta})\kappa_{\alpha_c}^\prime(\boldsymbol{\theta})+\frac{1}{2}(\bthetan-\boldsymbol{\theta})^2\kappa_{\alpha_c}^{\prime\prime}(\boldsymbol{\theta})\right\}+o\left(\frac{1}{n}\right)\\
		=&\text{E}_{\btheta}\left\{1-\alpha_c+(\bthetan-\boldsymbol{\theta})\kappa_{\alpha}^\prime(\boldsymbol{\theta})+\frac{1}{2}(\bthetan-\boldsymbol{\theta})^2\kappa_\alpha^{\prime\prime}(\boldsymbol{\theta})\right\}+O\left(\frac{1}{n^2}\right)+o\left(\frac{1}{n}\right)\\
		=&1-\alpha_c+c(\boldsymbol{\theta})/n+o\left(\frac{1}{n}\right)=1-\alpha+o\left(\frac{1}{n}\right),
	\end{split}
\end{equation*}
by expanding $\kappa_{\alpha_c}^\prime(\boldsymbol{\theta})$ and $\kappa_{\alpha_c}^{\prime\prime}(\boldsymbol{\theta})$ around $\alpha$.
In other words, the error rate of $1-\alpha$ plug-in prediction bounds can be improved from $O(1/n)$ to $o(1/n)$ by using an adjusted quantile $y_{1-\alpha_c}(\bthetan)$, rather than the $1-\alpha$ quantile $y_{1-\alpha}(\bthetan)$ directly, from plug-in cdf $G(\cdot;\bthetan)$ of $Y$.
A similar expansion can be obtained by replacing $\alpha_c$ with an estimator $\widehat{\alpha}_c=\alpha+c(\bthetan)/n$.
The drawback of this method is usually a closed form for $c(\cdot)$ in (\ref{eq:calibration-expansion}) is not available.

\subsubsection{Additional Comments}

In some special cases, $U_n$ does not converge to $\text{Uniform}(0,1)$ and has a limiting distribution function that depends on an unknown $\btheta$ (examples include \citealt{tian2020pred}).
The plug-in method then fails because using a $\text{Uniform}(0,1)$ to calibrate the distribution of $U_n$ is no longer valid, even asymptotically.
Nevertheless, we can still use the calibration-bootstrap method to construct asymptotically correct prediction intervals based on the non-pivotal quantity $U_n$.
In fact, we can use a broader notion of predictive root $q(\boldsymbol{X}_n,Y)$ (cf.~\citealt{beran1990}) to include both (approximate) pivotal and non-pivotal cases.
Although using a non-pivotal predictive root $q(\boldsymbol{X}_n,Y)$ usually leads to an asymptotically correct prediction interval, it does not have the benefit of being exact.

\section{The Predictive Distribution Concept}\label{sec:pred-dist-concept}

\subsection{Bayesian and Non-Bayesian Predictive Distributions}

The concept of a predictive distribution (free of unknown parameters) originated in Bayesian statistics, but efforts have been made to extend the predictive distribution idea to the non-Bayesian world.
Non-Bayesian predictive distributions have been implemented using terms including ``predictive distribution,'' ``predictive density,'' ``predictive likelihood,'' and ``prediction function.''
Although terminology varies, they have the same goal to ``express the relative credibility of the possible outcomes of a future experiment, in the light of a performed experiment'' (\citealt{mathiasen1979prediction}).

Although they may have similar forms, the Bayesian predictive distribution, however, is fundamentally different from these non-Bayesian predictive distributions.
The Bayesian predictive distribution always represents a type of conditional distribution of $Y$ given $\boldsymbol{X}_n=\boldsymbol{x}_n$, but this is not the case for the non-Bayesian predictive distribution.
When discussing non-Bayesian predictive distributions (in the form of a cdf), \citet{lawless2005} say ``In a loose sense these are conditional distributions $\tilde{F}_p(y|x)$ that provide probability statement about the future random variable $Y$, given $X=x$.
However, their properties are generally considered by treating them as estimators of the distribution function of $Y$ given $X=x$,'' where ``these'' means non-Bayesian predictive distributions in the form of a cdf.

Because our focus is on non-Bayesian methods and for the sake of simplicity, {\em we refer to the non-Bayesian predictive distributions as ``predictive distributions''} in the rest of the paper.
When discussing the Bayesian method, we still use the term ``Bayesian predictive distribution.''
The predictive distribution comes in three forms: the predictive cdf, pdf, and likelihood.
These three forms are closely related.
One can obtain a predictive cdf by integrating the corresponding predictive pdf and obtain a predictive pdf by normalizing the corresponding predictive likelihood.
We use $F_p(y;\boldsymbol{x}_n)$, $f_p(y;\boldsymbol{x}_n)$, and $L_p(y;\boldsymbol{x}_n)$, respectively, to denote the predictive cdf, pdf, and likelihood.
The notational purpose of using a semicolon instead of a vertical bar is because the predictive distribution is not the conditional distribution of $Y$ given $\boldsymbol{X}_n$.

\subsection{Prediction Interval Methods and Predictive Distributions}\label{subsec:interval-and-predictive-distribution}

For any prediction interval method, we can construct a predictive distribution for $Y$ by treating the endpoint of the $1-\alpha$ upper prediction bound as the $1-\alpha$ quantile of such distribution.
Specifically, there is a corresponding predictive distribution for any given prediction interval method.

For the pivotal and approximate pivotal calibration methods described in Section~\ref{sec:pred-interval}, \citet{lawless2005} give the formulas for finding the associated predictive cdf.
If a pivotal quantity $q(\boldsymbol{X}_n,Y)$ exists and has cdf $Q_n(\cdot)$, with the further assumption that $q(\boldsymbol{x}_n,y)$ is a monotone function of $y$, then the corresponding predictive cdf for the pivotal method based on $q(\boldsymbol{x}_n,y)$ is given by
$
F_p(y;\boldsymbol{x}_n)=Q_n[q(\boldsymbol{x}_n,y)]
$.
Similarly, if $q(\boldsymbol{X}_n,Y)$ is an approximate pivotal quantity with cdf $Q_n(\cdot;\btheta)$, the corresponding predictive cdf is $F_p(y;\boldsymbol{x}_n)=\widetilde{Q}_n[q(\boldsymbol{x}_n,y)]$, where $\widetilde{Q}_n(\cdot)$ is an estimate of $Q_n(\cdot;\btheta)$ (e.g., $\widetilde{Q}_n(\cdot)=\lim_{n\to\infty}Q_n(\cdot;\btheta)$).

For the calibration-bootstrap method, $H_n(\cdot;\bthetan)$ is used to approximate $H_n(\cdot;\btheta)$, where the latter is the cdf  of $U_n=G(Y|\boldsymbol{X}_n;\bthetan)$; thus the associated predictive cdf is
\begin{equation}\label{eq:calibration-bootstrap-pred}
	F_p(y;\boldsymbol{x}_n)=H[G(y|\boldsymbol{x}_n;\bthetan);\bthetan].
\end{equation}
When an explicit form of $H(\cdot;\cdot)$ is not available, \citet{fonseca2012} propose a formula to compute (\ref{eq:calibration-bootstrap-pred}) using bootstrap
\begin{equation}\label{eq:fonseca-formula}
\begin{split}
{F}_p(y;\boldsymbol{x}_n)&= \text{E}_{\bthetan}\bigg(G\left\{G^{-1}\left[G(y|\boldsymbol{x}_n; \widehat{\boldsymbol{\theta}}_{n})\big|X_{n}^\ast, \widehat{\boldsymbol{\theta}}_{n}^{\ast}\right]\big|X_{n}^\ast, \widehat{\boldsymbol{\theta}}_n\right\}\bigg)\\
&\approx\frac{1}{B}\sum_{b=1}^{B}G\left\{G^{-1}\left[G(y|\boldsymbol{x}_n; \widehat{\boldsymbol{\theta}}_{n})\big|\boldsymbol{x}_{n,b}^\ast, \widehat{\boldsymbol{\theta}}_{n,b}^{\ast}\right]\big|\boldsymbol{x}_{n,b}^\ast, \widehat{\boldsymbol{\theta}}_n\right\},
\end{split}
\end{equation}
where $\text{E}_{\bthetan}$ is the expectation with respect to the bootstrap sample $\boldsymbol{X}_n^\ast$ and the corresponding bootstrap estimate $\bthetan^\ast$; the second expression in (\ref{eq:fonseca-formula}) represents a Monte Carlo approximation based on the bootstrap estimates $\widehat{\boldsymbol{\theta}}_{n,b}^\ast$ from independently generated bootstrap samples $b=1,\dots,B$ for some $B$.
However, for values of $y$ where $G(y|\boldsymbol{x}_n;\bthetan)$ is close to one, the approximation formula in (\ref{eq:fonseca-formula}) will fail due to limited precision of floating-point computations.

\section{Predictive Distribution Methods}\label{sec:pred-dist-method}

\subsection{An Overview}

In Section~\ref{subsec:interval-and-predictive-distribution}, we describe finding the associated predictive distribution for prediction interval methods.
Conversely, given a predictive distribution, we can obtain the corresponding prediction intervals using the quantiles of that predictive distribution.
So, the development of a predictive distribution can be useful for formulating a prediction method.

\citet{bjo1990} summarizes three types of predictive likelihood methods (equivalently, predictive distribution methods): maximization-based, conditioning-based, and integration-based.
In addition to the methods discussed in \citet{bjo1990}, \citet{barncox1996} propose a predictive density that generally yields prediction intervals that have a coverage probability that is close to the nominal confidence level.
\citet{komaki1996asymptotic} considers constructing predictive distributions from the viewpoint of optimizing the Kullback-Leibler divergence between the true distribution of $Y$ and the predictive distribution of $Y$.
But this idea of constructing non-Bayesian predictive distribution is not without difficulty as \citet{hall1999} point out that many of the predictive distribution methods ``do not reduce coverage error by an order of magnitude, relative to the `naive' or `estimative' approach to prediction.''
They further use bootstrap calibration to improve the coverage.
In our review, we focus on the integration-based methods, where more research has been done since the review paper by \citet{bjo1990}.

\subsection{Integration-Based Predictive Distributions}
\label{subsec:integration-based-pred}

The construction of an integration-based predictive distribution is similar to that of the Bayesian predictive distribution in (\ref{eq:bayesian-predictive-distribution}).
The idea is to assign a data-based distribution to the non-random parameter $\btheta$ and use this distribution to marginalize out the parameters in the distribution function $G(y|\boldsymbol{x}_n;\btheta)$ of $Y$.
The resulting predictive cdf has the form
\begin{equation}\label{eq:integration-based-predictive-distributions}
F_p(y;\boldsymbol{x}_n)=\int G(y|\boldsymbol{x}_n;\btheta)p(\btheta;\boldsymbol{x}_n)d\btheta,
\end{equation}
where $p(\btheta;\boldsymbol{x}_n)$ is a data-based pdf assigned to $\btheta$.
More generally, we do not strictly require a pdf $p(\btheta;\boldsymbol{x}_n)$ for purposes of defining an integral of $G(y|\bm{x}_n;\bm{\theta})$ over $\bm{\theta}$ in (\ref{eq:integration-based-predictive-distributions}).
Technically, any data-based distribution over the parameter space can be used to integrate $G(y|\bm{x}_n;\bm{\theta})$ (although not all of them have a practical meaning) and, in practice, $F_p(y;\bm{x}_n)$ is often evaluated through a Monte Carlo approximation as
\[
F_p(y;\bm{x}_n) \approx \frac{1}{B} \sum_{b=1}^B G(y|\bm{x}_n;\bm{\theta}^{(b)})
\]
using a set of independent draws $\bm{\theta}^{(1)},\ldots,
\bm{\theta}^{(B)}$ from the chosen distribution over the parameter space
that is determined from the data 
$\bm{x}_n$.
In the rest of this section, we discuss three types of integration-based predictive distributions.

\subsubsection{Using a Bootstrap Distribution}
\label{subsubsec:bootstrap-distr}
\citet{harris1989} proposes a bootstrap predictive distribution obtained by integrating (\ref{eq:integration-based-predictive-distributions}) using a bootstrap distribution of $\bthetan$ in the role of $p(\btheta;\boldsymbol{x}_n)$ and shows that the proposed predictive distribution is asymptotically superior to the plug-in method in terms of average Kullback-Leibler divergence for the natural exponential family.
Although bootstrap samples are used, the coverage probability of this method can be shown, under assumptions similar to the plug-in method, to exhibit a coverage error of order $O(1/n)$, which is the same error rate as the plug-in method in (\ref{eq:calibration-expansion}); formal details are given in Section~A of the supplementary material.
We call this method the ``direct-bootstrap'' method because the bootstrap draws are used directly to compute the predictive distribution.
In Section~\ref{subsec:calibration-bootstrap-gpq}, we introduce the generalized pivotal quantity (GPQ) bootstrap method, a variant of this method.

\subsubsection{Using a Fiducial Distribution}
Fiducial inference was first introduced by R. A. Fisher and applies concepts of transferring randomness from the data to the parameters to produce a fiducial distribution on the parameter space.
The resulting fiducial distribution is similar to a Bayesian posterior but does not require a prior distribution.
We use an illustrative example to demonstrate the fiducial idea.
Suppose $X\sim\text{Norm}(\mu,1)$, then a structural equation for linking the data to the parameter is given by $X=\mu+Z$ where $Z\sim\text{Norm}(0,1)$.
For a realized $X=x$, this equation is solved for $\mu$ as $\mu=x-Z$ and thus the fiducial distribution for $\mu$ is $\text{Norm}(x, 1)$.
The construction of a fiducial distribution may not be unique.
More details about fiducial and generalized fiducial inference can be found in \citet{hannig2016generalized}.
When the data $\boldsymbol{X}_n$ and $Y$ are independent (i.e., $G(y|\boldsymbol{x}_n;\btheta)=G(y;\btheta)$), the fiducial predictive cdf has the same form as (\ref{eq:integration-based-predictive-distributions}), where $p(\btheta;\boldsymbol{x}_n)$ is the fiducial distribution of $\btheta$.
A detailed discussion of the fiducial prediction, including the case that $\boldsymbol{X}_n$ and $Y$ are dependent, can be found in \citet{wang2012}.

\subsubsection{Using a Confidence Distribution}

\citet{shen_liu_xie_2018} propose a prediction framework based on the notion of confidence distribution (CD) and prove that the corresponding prediction interval is asymptotically correct for a scalar parameter.
The idea is to replace $p(\btheta;\boldsymbol{x}_n)$ in (\ref{eq:integration-based-predictive-distributions}) with a real-valued confidence
distribution.
But as stated in \citet{xie_singh_2013}, the definition of confidence distribution for a parameter vector with more than one element remains an open question, and the theoretical properties of CD-based predictive distributions in this more general setting require further development.

\section{New Results for Location-Scale Distributions}\label{sec:location-scale}

This section presents some particular results for predicting an independent future random variable from a (log-)location-scale distribution given data from the same distribution.
These families of distribution include the most widely used probability distributions, such as the normal, lognormal, logistic, loglogistic, Weibull, Fr\'{e}chet, and some extreme value distributions.
Consider a sample $\boldsymbol{X}_n$ consisting of $n$ iid observations from a member of the location-scale distribution family with cdf $F(x;\mu,\sigma)=\Phi\left[(x-\mu) /\sigma\right]$ depending on parameters $\mu\in\mathbb{R}$ and $\sigma>0$ and where $\Phi(\cdot)$ is a given continuous cdf with no unknown parameters.
The corresponding pdf is then $f(x;\mu, \sigma)=\sigma^{-1}\phi[(y-\mu)/\sigma]$, where $\phi(z)=d\Phi(z)/dz$.
The predictand $Y$ is an independent random variable from the same distribution.
Suppose that the data $\boldsymbol{X}_n$ can be observed under three different situations: complete $\boldsymbol{X}_n$, time (Type-I) censored $\boldsymbol{X}_n^{\text{I}}$, or failure (Type-II) censored $\boldsymbol{X}_n^{\text{II}}$.
For time-to-event data, Type-I censoring means that observation stops at a fixed censoring time, while Type-II censoring means that
observation stops once a predetermined number of events have occurred.

\subsection{The Calibration-Bootstrap Method and its Predictive Distribution}
\label{subsec:calibration-bootstrap-gpq}

This section shows that (i) the calibration-bootstrap method (cf.~Section~\ref{subsubsec:calibration-bootstrap}) is equivalent to a predictive distribution based on integrating out the parameters with the distribution of the GPQ and (ii) the calibration-bootstrap method is also shown to be equivalent to a pivotal method (cf.~Section~\ref{subsec:pivotal-method}) for complete or Type-II censored data, thus having exact coverage probability.

By applying (\ref{eq:fonseca-formula}) to the location-scale distribution,
the predictive cdf of the calibration-bootstrap method is
\begin{equation}
	{F}_p(y;\boldsymbol{x}_n)=\text{E}{}_{\bthetan}\left(F\{F^{-1}\left[F(z;
	\widehat{\mu},\widehat{\sigma});\widehat\mu^\ast,\widehat\sigma^\ast
	\right];\widehat\mu,\widehat\sigma\}\right)=\text{E}{}_{\bthetan}\Phi\left\{\frac{y-[\widehat{\mu}+
		\frac{\widehat{\sigma}}{\widehat{\sigma}^{*}}(\widehat{\mu}-\widehat{\mu}^{*})]}
	{\widehat{\sigma}\frac{\widehat{\sigma}}{\widehat{\sigma}^{*}}}\right\};
	\label{eq:predictive-distribution-location-scale}
\end{equation}
here ($\widehat\mu$, $\widehat\sigma$) are the ML estimators of ($\mu, \sigma$) and $\text{E}_{\bthetan}$ denotes expectation with respect to the bootstrap distribution of $(\widehat{\mu}^\ast,\widehat{\sigma}^\ast)$, which is a version of $(\widehat{\mu},\widehat{\sigma})$ found from (parametric) bootstrap samples.

We define two new quantities
($\widehat\mu^{\ast\ast}, \widehat\sigma^{\ast\ast}$) using ($\widehat\mu, \widehat\sigma$) and ($\widehat\mu^\ast, \widehat\sigma^\ast$) as,
\begin{equation}\label{eq:gpqs-location-scale}
	\widehat\mu^{\ast\ast}\equiv\widehat{\mu}+\widehat{\sigma}\frac{\widehat{\mu}-\widehat{\mu}^{*}}{\widehat{\sigma}^{*}}
	,\quad
	\widehat\sigma^{\ast\ast}\equiv\widehat\sigma\frac{\widehat\sigma}{\widehat\sigma^{\ast}}.
\end{equation}
Then (\ref{eq:predictive-distribution-location-scale}) can be written in the form of (\ref{eq:integration-based-predictive-distributions}), where the parameters $(\mu, \sigma)$ in the cdf $F(y;\mu,\sigma)=\Phi\left[(y-\mu)/\sigma\right]$ of the predictand $Y$ are integrated out with respect to the joint distribution of $(\widehat\mu^{\ast\ast},\widehat\sigma^{\ast\ast})$ as
\begin{equation}\label{eq:gpq-predictive-distribution}
	{F}_p(y;\boldsymbol{x}_n)=\text{E}_{\bthetan}\Phi\left(\frac{y-\widehat\mu^{\ast\ast}}{\widehat\sigma^{\ast\ast}}\right)=\int\Phi\left(\frac{y-\widehat\mu^{\ast\ast}}{\widehat\sigma^{\ast\ast}}\right)\Pr{}_{\!\bthetan}(d\widehat{\mu}^{\ast\ast}, d\widehat{\sigma}^{\ast\ast})\approx\frac{1}{B}\sum_{b=1}^{B}\Phi\left(\frac{y-\widehat{\mu}^{\ast\ast}_b}{\widehat{\sigma}^{\ast\ast}_b}\right),
\end{equation}
where $(\widehat{\mu}^{\ast\ast}_b, \widehat\sigma^{\ast\ast}_b)$
are realized values of $(\widehat{\mu}^{\ast\ast}, \widehat{\sigma}^{\ast\ast})$ over independently
generated bootstrap samples $b=1,\dots,B$.
This equivalence shows that the calibration-bootstrap method
coincides with a predictive distribution constructed via an integration method.

Next, we introduce the definition of GPQ and illustrate the connection between the calibration-bootstrap method and GPQs.
Here we use the definition given in \citet{hannig2006}.
Let $\mathbb{S}\in\mathbb{R}^k$ denote a random vector and $\mathbb{S}^\ast$ is an independent copy of $\mathbb{S}$.
The distribution of $\mathbb{S}$ is indexed by $\btheta$.
Suppose we would like to estimate a function of $\btheta$ (possibly a vector) $\xi\equiv\pi(\btheta)$.
A GPQ for $\xi$, denoted by $\mathcal{R}_\xi$, is a function $(\mathbb{S},\mathbb{S}^\ast,\btheta)$ with the following properties
\begin{enumerate}[topsep=0pt,itemsep=-1ex,partopsep=1ex,parsep=1ex]
	\item The distribution of $\mathcal{R}_\xi$, conditional on $\mathbb{S}=\boldsymbol{s}$, is free of $\xi$.
	\item For every allowable $\boldsymbol{s}\in\mathbb{R}^k$, $\mathcal{R}_\xi$ depends on $\btheta$ only through $\xi$.
\end{enumerate}
We can use GPQs, for example, to construct confidence intervals for parameters of interest.

Interestingly, for complete data $\boldsymbol{X}_n$ or Type-II censored data
$\boldsymbol{X}_n^{\text{II}}$, the pair $(\widehat{\mu}^{\ast\ast}, \widehat{\sigma}^{\ast\ast})$ defined in (\ref{eq:gpqs-location-scale}) has the same
distribution as the GPQ $(\mu^{\ast\ast},\sigma^{\ast\ast})$, defined as
\begin{equation}
	\begin{aligned}
		\mu^{\ast\ast}=\widehat{\mu}+\left(\frac{\mu-\widehat{\mu}^{\mathbb{S}}}{\widehat{\sigma}^{\mathbb{S}}}\right) \widehat{\sigma}
		,\quad \sigma^{\ast\ast}=\left(\frac{\sigma}{\widehat{\sigma}^{\mathbb{S}}}\right) \widehat{\sigma},\end{aligned}
	\label{eq:gpq-definition}
\end{equation}
where $\mathbb{S}$ denotes an independent copy of the sample
$\boldsymbol{X}_n$ (or $\boldsymbol{X}_n^{\text{II}}$), and $(\widehat{\mu}, \widehat{\sigma})$ and $(\widehat{\mu}^{\mathbb{S}}, \widehat{\sigma}^{\mathbb{S}})$
denote the ML estimators of $(\mu, \sigma)$
computed from $\boldsymbol{X}_n$ (or $\boldsymbol{X}_n^{\text{II}}$) and
$\mathbb{S}$, respectively. The pair $(\mu^{\ast\ast},\sigma^{\ast\ast})$
is called the GPQ of $(\mu, \sigma)$ for location-scale distribution (cf. \citealt[Page 17]{krishnamoorthy2009statistical}).
Because $(\ref{eq:gpq-definition})$ are also fiducial quantities, (\ref{eq:predictive-distribution-location-scale}) is a fiducial predictive cdf (details are given in Section~B of the supplementary material).

Because the pair $(\widehat\mu^{\ast\ast},\widehat\sigma^{\ast\ast})$ is available for any (log-)location-scale distribution and because this pair is operationally computed from the bootstrap samples $(\widehat\mu^\ast, \widehat\sigma^\ast)$ (i.e., compare (\ref{eq:gpqs-location-scale}) to (\ref{eq:gpq-definition})), the prediction method in (\ref{eq:gpq-predictive-distribution}) is called the ``GPQ-bootstrap'' method in contrast to the ``direct-bootstrap'' method where the bootstrap pair $(\widehat\mu^\ast, \widehat\sigma^\ast)$ is used directly.
Note that under Type-I or random censoring, $(\widehat\mu^{\ast\ast},\widehat\sigma^{\ast\ast})$ are no longer GPQs; however, we can still use the prediction method in (\ref{eq:gpq-predictive-distribution}) the resulting prediction intervals are still asymptotically correct.

Note also that the calibration-bootstrap samples are used to approximate the quantity
\begin{equation*}
	U=G(Y;\widehat{\mu}, \widehat{\sigma})=\Phi\left(\frac{Y-\widehat{\mu}}{\widehat{\sigma}}\right)
	=\Phi\left[\frac{(Y-\mu)/\sigma-(\widehat{\mu}-\mu)/\sigma}{\widehat{\sigma}/\sigma}\right],
\end{equation*}
which is a pivotal quantity under complete or Type-II censored data
and its bootstrap re-creation also has the same distribution for such data.
This implies that, for complete or Type-II censored data, the calibration-bootstrap method has exact coverage probability and so does the GPQ-bootstrap method (i.e., due to producing the same prediction intervals from matching predictive distributions in (\ref{eq:predictive-distribution-location-scale}) and (\ref{eq:gpq-predictive-distribution})).
We provide illustrative numerical examples in Sections~E.1--4 of the supplementary materials.

\subsection{Properties of the Bayesian Predictive Distribution}
\label{subsec-properties-of-bayesian-pred}
For location-scale distributions and complete or Type-II censored data, the exact probability
matching prior is $\pi(\mu,\sigma)=\sigma^{-1}$ (and this is also known as the modified Jeffreys prior),
which implies that using this prior leads to credible intervals that
have exact frequentist coverage for either $\mu$ or $\sigma$
(cf. \citet{peers1965confidence}, \citet{lawless1972conditional}, \citet{diciccio2017}) and certain functions of these parameters (e.g., quantiles
and tail probabilities).
The purpose of this section is to show that (i) the prediction interval procedure based on the Bayesian predictive distribution using the prior $\pi(\mu,\sigma)=\sigma^{-1}$ is exact and (ii) the Bayesian predictive distribution using the prior $\pi(\mu, \sigma)=\sigma^{-1}$ is equivalent to a predictive distribution based on the generalized fiducial distribution (GFD) derived from the user-friendly formula in Section~2 of \citet{hannig2016generalized}.
The latter GFD for $(\mu,\sigma)$ has a density that is proportional to
\begin{equation*}
	r(\mu,\sigma;\boldsymbol{x}_n)\propto\frac{J(\boldsymbol{x}_n;\mu,\sigma)}{\sigma^n}\prod_{i=1}^{n}\phi\left(\frac{x_i-\mu}{\sigma}\right),
\end{equation*}
by Theorem~1 of \citet{hannig2016generalized}, where the function $J(\boldsymbol{x}_n;\mu,\sigma)$ is
\begin{equation*}
	J(\boldsymbol{x}_n;\mu,\sigma)=\sum_{1\leq i<j\leq n}\left|\textup{\textrm{det}}\left(\begin{bmatrix}
		1 & 1\\
		\frac{x_i-\mu}{\sigma} & \frac{x_j-\mu}{\sigma}
	\end{bmatrix}\right)\right|=\frac{1}{\sigma}\sum_{1\leq i<j\leq n}\left|x_i-x_j\right|.
\end{equation*}

\begin{theorem}
	Under a complete sample $\boldsymbol{X}_n$ or a Type-II censored sample $\boldsymbol{X}_n^{\text{II}}$
	from a location-scale distribution with location parameter $\mu$ and scale parameter
	$\sigma$, suppose that the ML estimators are $\widehat\mu$ and $\widehat\sigma$, $Y$ is an independent
	random variable from the same distribution as $\boldsymbol{X}_n$, and that the quantity $(U_1, U_2)$
	is defined as $((\widehat{\mu}-\mu)/\widehat{\sigma}, \widehat{\sigma}/\sigma)$. Then,
	\begin{enumerate}[topsep=0pt,itemsep=-1ex,partopsep=1ex,parsep=1ex]
		\item The joint posterior distribution of $(U_1, U_2)$ using prior
		$\pi(\mu,\sigma)\propto\sigma^{-1}$ is the same as the frequentist conditional distribution
		of $(U_1, U_2)$ conditioned on ancillary statistic $A=((X_1-\widehat\mu)/\widehat\sigma, \dots, (X_{n-2}-\widehat\mu)/\widehat\sigma)$.
		\item The $1-\alpha$ Bayesian upper prediction bound, which is defined as
		\begin{equation}\label{bayes-bound-eq}
			\widetilde{Y}_{1-\alpha}^{Bayes}\equiv\inf\left\{y: \int_{(\mu, \sigma)\in\boldsymbol{\Theta}}F(y|\mu, \sigma)p(\mu,\sigma|\boldsymbol{X}_n)d\mu d\sigma\geq1-\alpha\right\},
		\end{equation}
		has exact coverage probability, i.e., $\Pr\left(Y\leq\widetilde{Y}_{1-\alpha}^{Bayes}\right)=1-\alpha$, where $p(\mu, \sigma|\boldsymbol{X}_n=\boldsymbol{x}_n)$ is the joint posterior distribution using prior $\pi(\mu, \sigma)=\sigma^{-1}$.
		\item The GFD for $(\mu,\sigma)$ is the same as the Bayesian posterior distribution for $(\mu,\sigma)$ using the prior $\pi(\mu,\sigma)=\sigma^{-1}$, and application of this GFD in
		(\ref{eq:integration-based-predictive-distributions}) produces a predictive distribution
		$\int_{(\mu, \sigma)} F(y;\mu,\sigma)p(\mu,\sigma|\boldsymbol{X}_n)d\mu d\sigma$ and
		bounds $\widetilde{Y}_{1-\alpha}^{Bayes}$ that match the corresponding Bayesian analogs in (\ref{bayes-bound-eq}).
	\end{enumerate}
	\label{theo:bayes}
\end{theorem}
The proof of Theorem~\ref{theo:bayes} is provided in Section~C of the supplementary material.
Although the term ``fiducial'' is used both here and in Section~\ref{subsec:calibration-bootstrap-gpq} (in the context of $(\widehat{\mu}^{\ast\ast},\widehat{\sigma}^{\ast\ast})$ there),
the GFD for $(\mu,\sigma)$ in Point 3 of Theorem~\ref{theo:bayes}
is generally
different from (but close to) the distribution of the GPQ pair
$(\widehat\mu^{\ast\ast}, \widehat\sigma^{\ast\ast})$ from (\ref{eq:gpqs-location-scale}).
This is because the GPQs $(\widehat{\mu}^{\ast\ast}, \widehat{\sigma}^{\ast\ast})$
are based on the unconditional distribution of $(U_1, U_2)=((\widehat{\mu}-\mu)/\widehat{\sigma}, \widehat{\sigma}/\sigma)$ while GFD is determined
by the conditional distribution of $(U_1, U_2)$ given the ancillary statistics
$\boldsymbol{A}=(A_1,\dots,A_{n-2})$ (or $(A_1,\dots,A_{r-2})$ for Type-II censoring); the latter follows from Points 1 and 3 of Theorem~\ref{theo:bayes}.
The one exception is for the normal distribution, where the distribution
of the GPQ pair $(\widehat{\mu}^{\ast\ast},\widehat{\sigma}^{\ast\ast})$
will match the GFD for $(\mu, \sigma)$ (for which Basu's theorem gives
that $(U_1, U_2)$ is independent of $\boldsymbol{A}$).

\section{Other Continuous Distributions}
\label{sec:examples}

This section describes and illustrates prediction methods for two continuous distributions that are not in the (log-)location-scale family.

\subsection{The Gamma Distribution}\label{gamma-simulation}

The data $\boldsymbol{X}_n$ and the predictand $Y$ are independent samples
from a gamma distribution with pdf $f(x;\alpha,\lambda)=\lambda^\alpha x^{\alpha-1}\exp(-\lambda x)/\Gamma(\alpha)$.
A small-scale simulation study was done to compare:
(i) the plug-in method; (ii) the calibration-bootstrap method; (iii) the direct-bootstrap method; and (iv) the fiducial predictive distribution (cf.~Section~\ref{subsec:integration-based-pred}).
Because the gamma distribution does not belong to the (log-)location-scale
family, the GPQ-bootstrap is not applicable.
To implement methods (i), (ii), and (iii),
the ML estimates were computed using the \textbf{egamma} function
in R package \textbf{EnvStats}. For method (iv), two ways of
constructing the fiducial distribution were used.

The first is an approximate method proposed by \citet{chen2017approximate}.
From a gamma sample $\boldsymbol{X}_n$, define a scaled chi-square random variable $W(\alpha)$ as,
\begin{equation*}
	W(\alpha)\equiv2n\alpha\log\left(\frac{\bar{X}_n}{\prod_{i=1}^{n}X_i^{1/n}}\right)
	\sim c\chi^{2}_{v},
\end{equation*}
where $c$ and $v$ can be calculated as
$v=2\text{E}^2(W(\alpha))/\text{Var}(W(\alpha))$ and $c=\text{E}(W(\alpha))/v$.
Here the expectation and variance of $W(\alpha)$ are
\[\text{E}_{\btheta}(W(\alpha))=2n\alpha\text{E}(S_1),\quad\text{Var}_{\btheta}(W(\alpha))=4n^2\alpha^2\text{Var}_{\btheta}(S_1),\]
where $\text{E}(S_1)=-\log n+\psi(\alpha n)-\psi(\alpha)$, $\text{Var}(S_1)=-\psi_1(\alpha n)+\psi_1(\alpha)/n$, $\psi(\cdot)$ is
the digamma function, and $\psi_1(\cdot)$ is the trigamma function.
\citet{chen2017approximate}
suggested using a consistent estimator $\widehat\alpha$ of $\alpha$ to compute $\widehat{c}$ and $\widehat{v}$.
Then the approximate marginal fiducial distribution of $\alpha$ is
defined by the distribution of a quantity $\alpha_b$ where
\begin{equation}
	\alpha_b\sim\frac{\widehat{c}\chi_{\widehat{v}}^2}{2n\log\left(\frac{\bar{x}_n}{\prod_{i=1}^{n}x_i^{1/n}}\right)}.
	\label{k_marginal_fid}
\end{equation}
Given a fiducial draw $\alpha_b$ sampled as (\ref{k_marginal_fid}), the fiducial draw for $\lambda$, denoted by $\lambda_b$, can be sampled as $\chi^2_{2n\alpha_b}/(2\sum_{i=1}^{n}x_i)$ using the fact that $2\lambda\sum_{i=1}^{n}X_i\sim\chi^2_{2n\alpha}$. Then the fiducial pairs $(\alpha_b, \lambda_b)$, $b=1,\dots,B$ can be used to compute the fiducial predictive distribution for $Y$ via (\ref{eq:integration-based-predictive-distributions}).

The second approach is from \citet{wang2012} based on the user-friendly formula in \citet{hannig2016generalized}.
The fiducial distribution of $(\alpha,\lambda)$ is given by a density proportional
to
\begin{equation}
	\begin{split}
		r(\alpha,\lambda;\boldsymbol{x}_n)&\propto\frac{\lambda^{n\alpha-1}\exp[-\lambda\sum_{i=1}^{n}x_i+(\alpha-1)\sum_{i=1}^{n}\log x_i]}{\alpha\Gamma(\alpha)^n}\times\\
		\sum_{1\leq i<j\leq n}&x_i x_j\left|\frac{\Gamma(\alpha+1)\frac{\partial}{\partial\alpha}V_{\alpha,1}(\lambda x_i)}{(\lambda x_i)^\alpha\exp(-\lambda x_i)}-\frac{\Gamma(\alpha+1)\frac{\partial}{\partial\alpha}V_{\alpha,1}(\lambda x_j)}{(\lambda x_j)^\alpha\exp(-\lambda x_j)}\right|,
	\end{split}
	\label{gamma_fid}
\end{equation}
where $V_{\alpha,1}(\cdot)$ is the cdf of a $\text{Gamma}(\alpha, 1)$.
\citet{wang2012} used an importance sampling algorithm to generate
fiducial draws of $(\alpha,\lambda)$ from (\ref{gamma_fid}).

We used simulation to compare these methods mentioned above, and results are given in Section~E.5 of the supplementary material.
The calibration-bootstrap method has the best performance and the estimated coverage probability is close to the nominal confidence level, even when $n$ is small.
Two fiducial methods also have good coverage probabilities but not as good as the calibration-bootstrap method when $n$ is small.
The direct-bootstrap method has poor coverage probability and does not improve on the plug-in method.
As described in Section~\ref{subsubsec:bootstrap-distr}, it can be shown that prediction bounds from direct bootstrap often share a close correspondence to plug-in prediction bounds.
General theory, along with numerical illustrations for the gamma case, appear in Section~A of the supplementary materials.
Consequently, the plug-in and direct-bootstrap methods perform very similarly to each other, but not as well as the calibration-bootstrap approach.

\subsection{The Inverse Gaussian Distribution}\label{inv-gauss-dist}
The sample $\boldsymbol{X}_n$ and predictand $Y$ are independent
samples from an inverse Gaussian distribution with pdf
$$f(x;\mu,\lambda)=\sqrt{\frac{\lambda}{2\pi x^3}}\exp\left[-\frac{\lambda(x-\mu)^2}{2\mu^2x}\right].$$
As in Section~\ref{gamma-simulation}, a small scale simulation study was done to compare several methods:
(i) plug-in; (ii) calibration-bootstrap;
(iii) direct-bootstrap;
(iv) fiducial predictive distribution methods.

For methods (i),(ii), and (iii), the ML estimators are $\widehat\mu=\bar{X}_n$ and $\widehat\lambda=n/\sum_{i=1}^{n}(X_i^{-1}-\bar{X}_n^{-1})$.
For method (iv), \citet{najera2017fiducial} proposed a method to sample from the
fiducial distribution for $(\mu, \lambda)$. Because $\sum_{i=1}^{n}(X_i^{-1}-\bar{X}_n^{-1})\sim\chi^2_{n-1}/\lambda$, the marginal
fiducial distribution of $\lambda$ is given as
$$
\lambda\sim\frac{\chi^2_{n-1}}{\sum_{i=1}^{n}(x_i^{-1}-\bar{x}_n^{-1})}.
$$
Then given $\lambda_b$, which is sampled as above, a fiducial draw $\mu_b$ for $\mu$ can be obtained using the following steps:
\begin{enumerate}
	\item Generate $u_b$ from $\text{Uniform}(0, 1)$.
	\item Compute the quantile $q_{+\infty,\lambda_b}(u_b)\equiv\textbf{qinvgauss}(u_b, \mu=+\infty, \lambda = \lambda_b)$.
	\item If $\bar{x}_n/\lambda_b\geq q_{+\infty,\lambda_b}(u_b)$, $\mu_b=+\infty$. If $\bar{x}_n/\lambda_b< q_{+\infty,\lambda_b}(u_b)$, $\mu_b$ is obtained by solving the equation $\textbf{pinvgauss}(u_b, \mu_b, n)=\bar{x}_n/\lambda_b$.
\end{enumerate}
The inverse Gaussian quantile function \textbf{qinvgauss}
and cdf function \textbf{pinvgauss} are available in R package
\textbf{statmod}.

The inverse Gaussian simulation (given in Section~E.6 of the supplementary material) gives results that are similar to those for the gamma simulation.
Use of the direct-bootstrap method does not improve on the plug-in method in terms of coverage probability.
The fiducial method has good coverage probability but, as shown in the examples, sampling from a given fiducial distribution is often non-trivial.
Both theoretical results (cf.~Section~\ref{subsubsec:calibration-bootstrap}) and simulations have shown that the calibration-bootstrap method has the best coverage for several continuous distributions and it is also easy to implement.

\section{Prediction Methods for Discrete Distributions}\label{sec:discrete}

Previous sections in this paper considered prediction from a continuous distribution, and those methods can be applied to a wide variety of continuous distributions (e.g., see Section~E in the supplementary materials).
This section focuses on prediction methods for discrete distributions.
Following the two prediction principles discussed in the previous sections (i.e., (approximate) pivotal methods and  (Bayesian and non-Bayesian) predictive distribution methods), this section first discusses some general methods and then implements these methods for the binomial and Poisson distributions.
Additionally, we give some cautionary remarks on using the plug-in method for discrete distributions.
\subsection{Some General Methods}

\subsubsection{The Pivotal Conditional Cdf Method}
\label{sub-sec-piv-cdf-methods}
Let $\boldsymbol{X}_n$ be the data, $Y$ be the predictand, and $T(\boldsymbol{X}_n)$ be a statistic whose conditional distribution given a function of $\boldsymbol{X}_n$ and $Y$, say $R(\boldsymbol{X}_n, Y)$, is a discrete function that does not depend on any unknown parameters.
The pivotal conditional cdf method described in Section~\ref{subsubsec:pivotal-conditional-cdf-method} cannot be used directly because $G_{T|R}[T(\boldsymbol{X}_n)|R(\boldsymbol{X}_n, Y)]$ is no longer $\text{Uniform}(0,1)$ distributed.
Nevertheless, because $G_{T|R}[T(\boldsymbol{X}_n)|R(\boldsymbol{X}_n, Y)]$ is stochastically ordered with respect to the $\text{Uniform}(0,1)$ distribution (see Section~D of the supplementary material), the pivotal conditional cdf method can be extended to discrete distributions with slight modifications as long as $G_{T|R}[T(\boldsymbol{x}_n)|R(\boldsymbol{x}_n,y)]$ is a monotone function of $y$.
Without loss of generality, suppose that
$G_{T|R}[T(\boldsymbol{x}_n)|R(\boldsymbol{x}_n,y)]$ is a non-increasing function of $y$, the $1-\alpha$ lower and upper prediction
bounds are defined as
\begin{equation}\label{conservative-method}
	\undertilde{Y}_{1-\alpha}=\inf\left\{y:1-G_{T|R}\left[T(\boldsymbol{x}_n)-1|R(T(\boldsymbol{x}_n),y)\right]>\alpha\right\},\quad\widetilde{Y}_{1-\alpha}=\sup\left\{y:G_{T|R}\left[T(\boldsymbol{x}_n)|R(T(\boldsymbol{x}_n),y)\right]>\alpha\right\}.
\end{equation}
We call (\ref{conservative-method}) the conservative method because
the prediction bounds are guaranteed to have a coverage probability that
is greater or equal to the nominal confidence level (details are given in Section~D of the supplementary material).

There are other constructions of the pivotal (conditional) cdf method.
Suppose the conditional distribution of $Y$ given $R(\boldsymbol{X}_n,Y)$
does not depend on any parameters and the conditional cdf is $G[y|R(\boldsymbol{x}_n,y)]$. \citet{faulkenberry1973method} proposes the conditional method by defining the $1-\alpha$
lower and upper bounds as
\begin{equation}\label{conditional=method}
	\undertilde{Y}_{1-\alpha}=\sup\{y:G_{Y|R}[y-1|R(x,y)]\leq\alpha\},\quad
	\widetilde{Y}_{1-\alpha}=\inf\{y:G_{Y|R}[y|R(x,y)]\geq1-\alpha\}.
\end{equation}
As noted by \citet{dunsmore1976note}, however, the prediction bounds in (\ref{conditional=method}) may not exist in some situations (i.e., the set may be empty).

\subsubsection{Approximate Pivotal Methods}
\label{approximate-piv-method-discrete}
Similar to the idea in Section~\ref{subsec:approximate-pivotal-methods}, we can construct prediction intervals using approximate pivotal quantities.
Suppose $q(\boldsymbol{X}_n,Y,\btheta)\xrightarrow{d}U$, where $U$ does not depend on any parameters.
Then, if $\btheta$ is known, a $1-\alpha$ prediction interval (or bound) can be defined by
$
\{y:q(\boldsymbol{x}_n,y,\btheta)\leq u_{n,1-\alpha}\},
$
where $u_{n,1-\alpha}$ is the $1-\alpha$ quantile of $U$. When $\btheta$ is unknown one can replace $\btheta$ with a consistent estimator $\bthetan$, such as $\bthetan(\boldsymbol{X}_n)$, which is the ML estimator of $\btheta$ from the data $\boldsymbol{X}_n$.
Another choice is to use $\bthetan(\boldsymbol{X}_n,Y)$, which is the estimator from both the data and the predictand $(\boldsymbol{X}_n,Y)$.
After replacing $\boldsymbol{\theta}$ with an estimator $\bthetan$, we can construct a prediction interval for $Y$ by solving $
\{y:q(\boldsymbol{x}_n,y,\bthetan)\leq u_{n,1-\alpha}\},
$ for integer $y$,
as illustrated in Sections~\ref{binom_pred} and ~\ref{poi_pred}.

\subsubsection{Methods Based on Integration}

Using an objective prior (e.g., a Jeffreys prior), a Bayesian predictive
distribution can be used to construct prediction intervals, which may have good frequentist coverage probability, as illustrated in the rest of this section.
Similarly, the fiducial method also works when an (approximate) fiducial distribution is available.
The conditioning-based predictive likelihood (cf.~\citet{bjo1990}) can also be used, as illustrated in Sections~\ref{binom_pred} and \ref{poi_pred}.

\subsection{The Binomial Distribution}
\label{binom_pred}

Let $X\sim\text{Binom}(n, p)$ and $Y\sim\text{Binom}(m, p)$, where $p\in(0, 1)$ is unknown and $n, m$ are given positive integers.
The goal is to construct prediction bounds for $Y$ based on the observed value of $X=x$. When $X=0$ or $X=n$, the ML estimate is $\widehat{p}=0$ or $\widehat{p}=1$ so that prediction methods based on ML estimators (including plug-in, calibration-bootstrap, and direct-bootstrap methods) cannot be
used directly;
this is because estimated distributions used for prediction are degenerate for
the extreme values of $X$.
Several prediction methods for the binomial case are described below.
A numerical study was done to compare some of the methods and the results are given in Section~E of the supplementary material.

\subsubsection{The Conservative Method}
\citet{thatcher1964relationships} notes that a prediction interval can be obtained by using the conditional cdf of $X$ given $X+Y$ and proposes this method, which is an implementation of the method described in Section~\ref{sub-sec-piv-cdf-methods}. Suppose there are $n+m$ balls and $R=X+Y$ are red balls. Then $X$ is the number of red balls out of $n$ balls,
which has a hypergeometric distribution $\text{Hyper}(X+Y, n, n+m)$ with cdf $\text{phyper}(\cdot; X+Y, n, n+m)$.
After observing $X=x$, the $1-\alpha$ lower and upper prediction bounds using the conservative method are
\begin{equation*}
		\undertilde{Y}_{1-\alpha}=\inf\left\{y:1-\text{phyper}(x-1;x+y,n,n+m)>\alpha\right\},\quad\widetilde{Y}_{1-\alpha}=\sup\left\{y:\text{phyper}(x;x+y,n,n+m)>\alpha\right\}.
\end{equation*}

\subsubsection{Methods based on Approximate Pivots}
The methods discussed in this section are implementations of
the general method described in Section~\ref{approximate-piv-method-discrete}.
By the Central Limit Theorem (CLT),
both $X$ and $Y$ have normal limits (as $m,n\to\infty$) in that
\begin{equation*}
	Z_X=\frac{X/n-p}{\sqrt{(1-p)p/n}}\xrightarrow{d}\text{Norm}(0, 1),\quad
	Z_Y=\frac{Y/m-p}{\sqrt{(1-p)p/m}}\xrightarrow{d}\text{Norm}(0, 1).
\end{equation*}
Because $X$ and $Y$ are independent and by standardizing $\sqrt{n}Z_Y-\sqrt{m}Z_X$ (as approximately $\text{Norm}(0,n+m)$), it also holds (as $m,n\to\infty$) that
\begin{equation}\label{piv-binom-asymp}
	\frac{ \sqrt{n}Z_Y-\sqrt{m}Z_X}
	{\sqrt{\text{Var}(\sqrt{n}Z_Y-\sqrt{m}Z_X)}}=
	\frac{Y-mX/n}{\sqrt{(n+m)({m}/{n})p(1-p)}}\xrightarrow{d}\text{Norm}(0, 1).
\end{equation}
As mentioned in Section~\ref{approximate-piv-method-discrete}, we need
to replace the parameter ($p$ above) with an estimator to construct prediction intervals.

\citet{nelsonapplied} proposes replacing $p$ in (\ref{piv-binom-asymp}) with $\widehat{p}_x=X/n$. But numerical studies in \citet{wang2008coverage} and \citet{krishnamoorthy2011improved} show that this method has poor coverage
probability, even for large samples. Instead
of using $\widehat{p}_x$, \citet{krishnamoorthy2011improved} propose replacing $p$ with
$\widehat{p}_{xy}=(X+Y)/(n+m)$ (along with a continuity adjustment: if $x=0$, use $x=0.5$; if $x=n$, use $x=n-0.5$). Inspired by
the Wilson score confidence interval (cf.~\citet{wilson1927probable}), \citet{wang2010closed} proposes another method where $\widehat{p}=(X+Y+z^2_{1-\alpha}/2)/(n+m+z^2_{1-\alpha})$
is used to replace $p$ in (\ref{piv-binom-asymp}).

\subsubsection{Methods Based on Integration}
For the $\mathrm{Binom}(n, p)$,
the Jeffreys prior is $\pi_{J}(p)\propto p^{-1/2}(1-p)^{-1/2}$
(i.e., $\pi_{J}(p)\sim\mathrm{Beta}(0.5, 0.5)$).
The $1-\alpha$ lower and upper Jeffreys (Bayesian) prediction bounds are
\begin{equation*}
	\utilde{Y}_{1-\alpha}=\mathrm{qbetabinom}(\alpha; m, x+0.5, n-x+0.5),\quad
	\widetilde{Y}_{1-\alpha}=\mathrm{qbetabinom}(1-\alpha;m, x+0.5, n-x+0.5),
\end{equation*}
where $\text{qbetabinom}(p;n, a, b)$ is the $p$ quantile of the beta-binomial
distribution with a sample-size parameter $n$ and shape parameters $a$ and $b$.
Compared with the method proposed by \citet{krishnamoorthy2011improved}, this Jeffreys prediction method is slightly more conservative (cf.~\citealt[Chapter 6]{meeker2017statistical}).

A fiducial quantity for parameter $p$ based on observation $X=x$ has the form
\begin{equation*}
	\mathcal{R}_p=U_{(x)}+D(U_{(x+1)}-U_{(x)}),
\end{equation*}
where $U_{(x)}$ is the $x$th smallest value out of $n$ independent $\text{Uniform}(0, 1)$ random variables ($U_{(0)}=0$, $U_{(n+1)}=1$) and $D\sim\mathrm{Uniform}(0,1)$. Integrating out the parameter $p$ in the $\text{Binom}(m,p)$ cdf using the density function of $\mathcal{R}_p$, say $r(p|X=x)$, gives the fiducial predictive distribution for $Y$ via (\ref{eq:integration-based-predictive-distributions}).
The prediction bounds are defined using the appropriate quantiles of the fiducial predictive distribution.

\subsubsection{The Hinkley Predictive Likelihood}

\citet{hinkley1979predictive} proposes a conditioning-based predictive likelihood, which is based on the fact that the conditional distribution of $X$ (or $Y$) given $X+Y$ has a hypergeometric distribution.
However, unlike the pivotal conditional cdf method (cf.~Section~\ref{sub-sec-piv-cdf-methods}), this method is invariant to whether the conditional cdf of $X$ or $Y$ is chosen because the predictive likelihood of $Y$ is
$$
{L}_p(y;x) = \frac{\binom{n}{x}\binom{m}{y}}{\binom{n+m}{x+y}},\quad y\in\{0,\dots,m\},
$$
where $x$ and $y$ are interchangeable.
The predictive cdf is obtained by normalizing the predictive likelihood as
$
F_p(y;x)={\sum_{j=0}^{y}{L}_p(j;x)}/{\sum_{i=0}^{m}{L}_p(i;x)}.
$
The $1-\alpha$ lower and upper prediction bounds are defined as
\begin{align*}
	\undertilde{Y}_{1-\alpha}=\sup\{y:{F}_p(y-1;x)\leq\alpha\},\quad
	\widetilde{Y}_{1-\alpha}=\inf\{y:{F}_p(y;x)\geq1-\alpha\}.
\end{align*}
Although both the conservative method in (\ref{conservative-method}) and the Hinkley predictive likelihood method are based on the conditional distribution of $X$ given $X+Y$, they lead to different prediction intervals.

\subsection{The Poisson Distribution}\label{poi_pred}
Suppose $X\sim\text{Poi}(n\lambda)$ and $Y\sim\text{Poi}(m\lambda)$, where $\lambda>0$ is unknown, $n$ and $m$ are known positive real values, and $Y$ is independent of $X$. The goal is to construct a prediction interval for $Y$ based
on the observation $X=x$.
Methods similar to those used in Section~\ref{binom_pred} are used for Poisson prediction.
\subsubsection{The Conservative Method}

Because the conditional distribution of $X$ given $X+Y$ is $\text{Binom}(x+y, n/(n+m))$, we can use the general method described in Section~\ref{sub-sec-piv-cdf-methods}.
The $1-\alpha$ lower and upper prediction bounds using the conservative method are
\begin{equation*}
	\begin{split}
		\undertilde{Y}_{1-\alpha}&=\inf\left\{y:1-\text{pbinom}(x-1;x+y,n/(n+m))>\alpha\right\},\\		\widetilde{Y}_{1-\alpha}&=\sup\left\{y:\text{pbinom}(x;x+y,n/(n+m))>\alpha\right\}.
	\end{split}
\end{equation*}
\subsubsection{Methods based on Approximate Pivots}
This section implements the methods proposed in Section~\ref{approximate-piv-method-discrete}.
By the CLT, both $X$ and $Y$ have normal limits (as $m,n\to\infty$) given by
\begin{equation*}
	Z_X=\frac{X-n\lambda}{\sqrt{n\lambda}}\xrightarrow{d}\text{Norm}(0, 1),\quad
	Z_Y=\frac{Y-m\lambda}{\sqrt{m\lambda}}\xrightarrow{d}\text{Norm}(0, 1).
\end{equation*}
Because $X$ and $Y$ are independent, $\sqrt{n}Z_Y-\sqrt{m}Z_X$ has approximately a normal distribution with mean 0 and variance $(n+m)$.
Thus, it holds (as $m,n\to\infty$) that
\begin{equation*}
	\frac{\sqrt{n}Z_Y-\sqrt{m}Z_X}{\sqrt{\text{Var}(\sqrt{n}Z_Y-\sqrt{m}Z_X)}}=
	\frac{Y-mX/n}{\sqrt{(m+m^2/n)\lambda}}\xrightarrow{d}\text{Norm}(0,1).
\end{equation*}
\citet{nelsonapplied} replaces the unknown $\lambda$ with
$\widehat\lambda_x=X/n$ and \citet{krishnamoorthy2011improved}
replace $\lambda$ with $\widehat\lambda_{xy}=(X+Y)/(n+m)$ (along with a
continuity adjustment: if $x=0$, use $x=0.5$).
Krishnamoorthy and Peng show that their method has
better coverage probability properties than Nelson's method.

\subsubsection{Methods Based on Integration}
The Jeffreys prior for the Poisson rate parameter is $\pi_J(\lambda)=\sqrt{1/\lambda}$, $\lambda>0$, and the corresponding posterior distribution is gamma(x+1/2, n) with density
$
p(\lambda|x)\propto\lambda^{x-1/2}\exp(-n\lambda)$, for $\lambda>0
$.
Using this posterior, the Bayesian predictive density for the Poisson distribution is
\begin{equation}\label{pois-bayes}
	p(y|x) = \frac{\Gamma(y+x+1/2)}{\Gamma(x+1/2)\Gamma(y+1)}
	\left(\frac{n}{n+m}\right)^{x+1/2}\left(1-\frac{n}{n+m}\right)^y,
\end{equation}
which is a negative-binomial distribution $\text{NB}(x+0.5, n/(n+m))$.
The Jeffreys Bayesian prediction method tends to be more conservative than
the method proposed by \citet{krishnamoorthy2011improved}, especially
when the ratio $m/n$ is small (i.e., the expected value of $n\lambda$ of the data $X$ greatly exceeds that $\lambda m$ of the predictand $Y$), but is less conservative than the conservative method (cf.~\citealt[Chapter 7]{meeker2017statistical}).

An approximate fiducial quantity for $\lambda$ given observation $X=x$ has
a distribution of a scaled chi-square variable $\chi^2_{2x+1}/2n$ (cf.~\citealt{dempster2008dempster}, \citealt{krishnamoorthy2010inference}).
Using this approximate fiducial distribution in place of the (gamma) posterior
$p(\lambda|x)$ in (\ref{pois-bayes}) leads to a fiducial predictive distribution for a Poisson predictand.

\subsubsection{The Hinkley Predictive Likelihood}
Using the fact that the conditional distribution of $X$ given $X+Y$
has a binomial distribution,
the Hinkley predictive likelihood is as follows
$$
{L}_p(y;x)=\frac{f(X=x, Y=y)}{f(X+Y=x+y)}=\frac{(x+y)!}{x!y!}\left(\frac{m}{n+m}\right)^y\left(\frac{n}{n+m}\right)^x,
y\geq0.
$$
Often, the predictive distribution obtained by normalizing the
predictive likelihood has no closed form, but in this example the
predictive pmf for $Y$
induced by ${L}_p(y;x)$ has a negative-binomial mass function
\[
f_p(y;x)=f_p(y;x+1,n/(n+m))=\binom{x+y}{x}[m/(n+m)]^{x}[n/(n+m)]^{y},\quad y=0,1,2,\dots
\]
Thus, the prediction bounds can be obtained by using the appropriate quantiles of the
negative-binomial distribution.

\subsection{Cautionary Comments about the Plug-in Method}
\label{sec-caution-plug-in}
The plug-in method generally works only when the following holds
\begin{equation}
	\label{uniform-convergence-1}
	\sup_{y\in\mathbb{R}}|G(y|\boldsymbol{X}_n; \btheta_0)-G(y|\boldsymbol{X}_n; \bthetan)|\xrightarrow{p}0
\end{equation}
as $n\to\infty$.
Here $G(\cdot|\boldsymbol{x}_n;\btheta_0)$ is the conditional distribution function of $Y$ given $\boldsymbol{X}_n$.
The convergence (\ref{uniform-convergence-1}) implies that the ``plug-in'' version of the cdf approaches the true cdf as $n\to\infty$.
This convergence, however, does not always hold.

\citet{tian2020pred} discuss a particular type of within-sample prediction problem, where a Type-I censored time-to-event dataset is given to predict the number of future events during a time period after the censoring time.
They show that (\ref{uniform-convergence-1}) does not hold in this within-sample prediction problem.
Thus, for this situation, the plug-in method is {\em not} asymptotically correct, which means, no matter how large the sample size $n$ is, the true coverage probability is generally different from the nominal confidence level.
Note that, for the within-sample prediction, the sample size of the data is $n$ while the scalar predictand is a summary statistic of $n-r$ Bernoulli random variables, where $r$ is the number of events observed in the data.
The plug-in method is invalid for a case where the predictand sample size $n-r$ is potentially large compared to the data sample size $n$.
Alternatively, \citet{tian2020pred} propose three bootstrap-based methods that are asymptotically correct.

Relatedly, for the discrete new-sample prediction problems in Sections~\ref{binom_pred} and \ref{poi_pred}, both the data $X$ and the predictand $Y$ may be generally viewed as counts that summarize two samples: one sample of size $n$ for $X$ and another sample of size $m$ for $Y$.
We emphasize that the plug-in prediction method requires cautious consideration of the relative sizes of $n$ and $m$.
Importantly, the plug-in method will similarly fail for the binomial and Poisson prediction cases of Sections~\ref{binom_pred} and \ref{poi_pred} more broadly, unless $m$ is appropriately small relatively to $n$ (that is, asymptotically, success for the plug-in method requires $m/n\to0$ implying that the data sample size $n$ dominates the predictand sample size $m$).
Without this condition, the $\text{Binom}(m,p)$ or $\text{Poi}(m\lambda)$ distribution of a predictand $Y$ cannot be consistently approximated by the plug-in prediction method (i.e., by substituting $X/n$ for $p$ or $\lambda$).
For this reason, plug-in prediction is not included in Section~\ref{binom_pred} or \ref{poi_pred}.
The other prediction methods in these sections, however, are valid without restrictions on the relative sizes of $m,n$.
That is, for any ratio $m/n$, these other prediction methods are asymptotically correct.

\section{Overview of Prediction Methods for Dependent Data}
\label{sec:dependent-data}

In this section, ``dependent data'' refers to data with a complicated dependence structure.
Common examples include time series, spatial data, and random networks, where the strength of dependence among observations often depends on proximity.
Other examples include mixed-effects models, longitudinal data, and small area estimation where correlation exists among observations sharing random effects or repeated measures.
In such examples, distributional models for predictands often share non-trivial connections to data models through dependence conditions.
See \citet{clarke_clarke_2018} for descriptions of other prediction applications with dependent data.

While the literature for predictions with independent data is more extensive, similar prediction methods exist for dependent data along the lines of the plug-in, calibration-bootstrap, and (integration-based) predictive distribution methods.
This section discusses prediction interval methods for dependent data.

Similar to previous sections, the prediction problems considered here are based on a parametric model for the data and the predictand, although the dependence among and between these quantities can create complications.
Providing a challenge with model formulation and prediction under dependence, both the data structure and the nature of possible dependence in samples can vary greatly.

A further challenge in prediction with dependent data is that prediction strategies developed for independent observations may fail if directly applied to dependent data, so that caution may be required.
As a simple example, the plug-in method provides consistent prediction bounds for many problems with independent data but fails for the within-sample prediction problem described in Section~\ref{sec-caution-plug-in} where, despite arising from a random sample, the predictand and the observed data are dependent.
Additionally, several prediction methods from previous sections involve bootstrap sampling (under independence), where data simulation and the generation of bootstrap samples is more complicated when data are dependent.

Before describing prediction methods, we mention that there exists a variety of ways to generate bootstrap samples under dependence, particularly for time series.
For the latter, common formulations of bootstrap include model-residual-based bootstraps (e.g., AR(p) models, cf.~\citealt{pan2016bootstrap}), transformation-based bootstraps aiming to weaken dependence (e.g.,  \citealt{kreiss2011range}, \citealt{jentsch2015covariance}),  and block-based bootstraps that use data blocks to reconstruct time series (\citealt{gregory2018smooth}).
These bootstrap methods differ in their mechanics as well as in the amount of time series structure presumed by the bootstrap.
For reviews of bootstrap methods with time series and other dependent data, see \citet{politis2003impact}, \citet{lahiri2006bootstrap}, and \citet{kreiss2012bootstrap}.

For parametric-model based predictions from dependent data, Sections~\ref{plug-in-dependent-data}, \ref{calibration-method-dependent-data}, and \ref{subsec-pred-dist-integration} respectively describe plug-in, calibration-bootstrap, and integration-based predictive methods.
The bootstrap, when employed, is parametric.
These procedures have been largely studied and justified in the context of Gaussian process models, where Gaussian assumptions also facilitate generation of bootstrap samples needed for the calibration-bootstrap.
More development is needed to extend these approaches to predictions with non-Gaussian dependent data, with some possibilities suggested in Section~\ref{subsec-non-gaussian}.

\subsection{The Plug-in Method}\label{plug-in-dependent-data}

\citet{beran1990} and \citet{hall1999} consider the plug-in prediction method for some specially structured dependent data
models with independent additive errors (e.g., regression models, and the AR(1) model).
To set an $h$-step ahead prediction interval, given a realization of time series data from an ARMA process, 
for example, \citet[Chapter 3]{brockwell2016introduction} suggest using a normality assumption along with an approximation for the best linear predictor found by replacing
unknown parameters with consistent estimates. Similarly, by assuming a stationary Gaussian process, the plug-in prediction interval has been suggested in
spatial applications based on using a normal approximation with kriging predictors, where unknown
parameters are replaced with consistent estimates (cf.~\citealt[Chapter 3]{cressie2015statistics}). With such a Gaussian process, the coverage probability of the plug-in method typically
has an error of $O(1/n)$ (cf.~\citealt{sjostedt2003bootstrap}, \citealt{vidoni2004improved}).
However, it is not generally clear when the plug-in method is asymptotically correct for dependent data, particularly for more complicated and potentially non-Gaussian dependence structures.

\subsection{The Calibration Method}\label{calibration-method-dependent-data}

Using the calibration-bootstrap method (described in Section~\ref{subsubsec:calibration-bootstrap}), \citet{sjostedt2003bootstrap} improve plug-in kriging prediction intervals for a stationary Gaussian process while \citet{de2015prediction} establish similar findings for predicting spatial averages from these processes, and \citet{hall2006parametric} use calibration-bootstrap for small area prediction.
Similar to the method described in Section~\ref{subsec:calibration-expansion}, \citet{giummole2010improved} calibrate the plug-in method using an asymptotic expansion for a general class of Gaussian models, including time-series, Gaussian state-space models, and Gaussian Markov random fields.
Under regularity conditions with certain dependent Gaussian processes, the calibration methods reduce the error of the coverage probability to $o(1/n)$ compared to $O(1/n)$ for the plug-in method (cf.~\citealt{sjostedt2003bootstrap}, \citealt{giummole2010improved}).

\subsection{Bayesian and Fiducial Predictive Distributions}
\label{subsec-pred-dist-integration}
The fiducial method can be potentially extended to dependent data, but requires the development of an appropriate fiducial distribution under dependence, as described in \citet{wang2012}.
The Bayesian method has been studied extensively for dependent data for prediction.
For example, \citet{west2006bayesian} discuss Bayesian prediction in dynamic linear models for time series; recent work includes \citet{aktekin2018sequential}, \citet{mcalinn2019dynamic}, and \citet{berry2020bayesian}.
\citet{handcock1993bayesian} propose a best linear unbiased prediction procedure within a Bayesian framework for Gaussian random fields and \citet{de1997bayesian} present prediction methods for some types of non-Gaussian random fields.
See \citet{hulting1991some}, \citet{harville1992classical}, and \citet{christensen2002bayesian} for related results in the context of linear mixed models or generalized mixed models.

\subsection{Extensions to Non-Gaussian Dependent Data}\label{subsec-non-gaussian}
As described in Sections~\ref{plug-in-dependent-data}-\ref{calibration-method-dependent-data}, many of the existing formal treatments of model-based predictions for dependent data have largely focused on types of Gaussian  processes.
This simply indicates that the prediction methods based on the plug-in or bootstrap methods rely heavily on tractable forms for the distribution of the dependent data.
One option for model-based predictions with non-Gaussian data is to use a Gaussian model in conjunction with a suitable data transformation; for example, \citet{de2009shortest} develop plug-in and bootstrap calibration for log-Gaussian fields.
Beyond normal data cases, we mention another general model class for developing predictions could potentially involve Markov random field (MRF) structures.
This approach for modeling dependent data involves specifying a full conditional distribution for each observation on the basis of an underlying MRF (cf.~\citet{besag1974spatial}).
Model formulation in a conditional, component-wise fashion provides an alternative to direct specification of a full joint distribution for the data.
Additionally, such conditional distributions often depend functionally on small subsets of ``neighboring" observations, which is a property that may be useful for extending the plug-in and calibration-bootstrap prediction methods to MRF models.
The supplement describes more details about this prediction problem, along with a numerical illustration (see Section~F of the supplementary materials).
For implementing parametric bootstrap  without assumptions of Gaussianity, an attractive feature of MRF models is that data may be simulated rapidly from specified full conditional distributions through Gibbs sampling (\citet{kaplan2020simulating}). 
Note that MRF models have applications to both continuous and discrete dependent data (cf.~\citealt{cressie2015statistics},~\citealt{kaiser2000construction}, and~\citealt{casleton2017local}).

Regarding the calibration-bootstrap of  Section~\ref{calibration-method-dependent-data}, the bootstrap for dependent data can be  difficult to  establish through the prescription in Section~\ref{subsec:approximate-pivotal-methods},  which  requires an analytic form for the conditional  distribution of a predictand $Y$ given the data $\bm{X}_n$.  
By ignoring this distribution, an alternative strategy for predictions based on bootstrap is to approximate the distribution of a prediction error $|Y-\hat{Y}|$, where $\hat{Y}$ denotes a statistic based on data $\bm{X}_n$.
See, for example, \citet{politis2013model} and \citet{pan2016bootstrap} for illustrations with time series, and also \citet{de2015jsm} for similar bootstrap predictions with spatial data.

\section{Nonparametric Prediction Methods}\label{sec:model-free}

Up to this point, we have considered prediction problems based on parametric models.
Given a sufficient amount of data, however, nonparametric methods are available to construct prediction intervals.
In this section, we discuss two types of nonparametric prediction methods.

\subsection{Prediction Intervals Based on Order Statistics}

Let $X_1,\dots,X_n$ be a random sample from some continuous distribution function $F(x)$, then $(X_{(r)}, X_{(s)})$ is a $100\left[(s-r)/(n+1)\right]\%$ prediction interval for an independent future random variable $Y$ from the same distribution, where $1\leq r<s\leq n$.
The coverage probability of this prediction interval method is exact and it does not depend on the form of the underlying continuous distribution (i.e., the method is distribution-free).
For a desired nominal coverage probability that cannot be obtained in the form of $[(s-r)/(n+1)]$, \citet{beran1993interpolated} suggest interpolation to construct nonparametric prediction intervals to approximate the desired coverage probability.
Also proposing methods based on order statistics, \citet{fligner1976some} consider constructing distribution-free
prediction intervals that contain at least $k$ of $m$ future random variables
and prediction intervals for any particular order statistic of a future sample.
\citet[Chapter 5 and Appendix G]{meeker2017statistical}
describe computational details and illustrate the use of these distribution-free methods.
\citet{frey2013data} proposes a shortest nonparametric prediction interval,
as opposed to methods that have approximately equal one-sided coverage probabilities.

\subsection{Conformal Prediction}
\label{subsec:conformal-prediction}
Conformal prediction has been gaining popularity recently because it applies to many prediction problems in the area of supervised learning, including regression and classification problems.
Conformal prediction has often been presented in the form of an algorithm (e.g., \citealt{vovk2005algorithmic}, \citealt{shafer2008tutorial}).
Here we describe the conformal prediction through sampling distributions in order to connect with the pivotal cdf methods described in Section~\ref{subsubsec:pivotal-conditional-cdf-method}.
%

Suppose the data sample $\boldsymbol{X}_n=\left\{X_1,\dots,X_n\right\}$ and the predictand $Y\equiv X_{n+1}$ are i.i.d. (or a weaker exchangeable assumption).
Conformal prediction intervals are based on a choice of distance statistic $d(\boldsymbol{X}_n,Y)$ (or nonconformity measure).
One example is $d(\boldsymbol{X}_n,Y)=|\bar{X}_n-Y|$, which is the distance between a new observation $Y$ and the data sample mean $\bar{X}_n$.
Denote the cdf of $d(\boldsymbol{X}_n,Y)$ by $G(t)=\Pr[d(\boldsymbol{X}_n, Y)\leq t],t\in\mathbb{R}$, and write the left limit of the cdf as $G(t-)=\Pr[d(\boldsymbol{X}_n, Y)< t]=\lim_{x\uparrow t}G(t),t\in\mathbb{R}$.
The conformal prediction uses the probability integral transform in combination with the quantile $1-\alpha$ of a $\text{Uniform}(0,1)$ (i.e., $1-\alpha$) to calibrate prediction regions (similar to the approach used in Section~\ref{subsubsec:pivotal-conditional-cdf-method}), after an initial step of estimating the cdf $G$ with an empirical distribution $\widehat{G}_y$.

Let $\boldsymbol{X}_{n+1}\equiv\left\{X_1,\dots,X_n,X_{n+1}\right\}$.
Then a $1-\alpha$ conformal prediction region for $Y$ is given as
\begin{equation}\label{eqn:con}
	\left\{y \in \mathbb{R}: \widehat{G}_{y}\left[d\left(\boldsymbol{X}_{n}, y\right)-\right]<1-\alpha\right\},
\end{equation}
where $\widehat{G}_y$ is given by
$$
\widehat{G}_{Y}(t)=\frac{1}{n+1} \sum_{i=1}^{n+1} \text{I}\left(d\left(\boldsymbol{X}_{n+1} \backslash\left\{X_{i}\right\}, X_{i}\right) \leq t\right), \quad t \in \mathbb{R}
$$
which is the empirical cdf of distances $d(\boldsymbol{X}_{n+1}\backslash\{X_i\},X_i),i=1,\dots,n+1$, found by separating point $X_i$ from $\boldsymbol{X}_{n+1}$.
Note that $\widehat{G}_{Y}(\cdot)$ depends on the unobserved value $X_{n+1}=Y$, so that $\widehat{G}_y(\cdot)\equiv\widehat{G}_{Y=y}(\cdot)$ is computed provisionally using a potential value of $Y=y$ in (\ref{eqn:con}).
Furthermore, in (\ref{eqn:con}),  the estimated cdf $\widehat{G}_y(t)$ at $t=d(\boldsymbol{X}_n, y)$  is technically replaced with a left limit $\widehat{G}_y(d(\boldsymbol{X}_n, y)-)$ as a type of adjustment to  $\widehat{G}_y$  (i.e., as the latter  always jumps by $1/(n+1)$ at the argument   $t=d(\boldsymbol{X}_n, y)$).
For perspective,  in other prediction problems based on cdf transforms, a switch from cdfs to left limits of cdfs is known to be helpful for correcting  issues of discreteness in data, with the effect of ensuring  coverage probabilities are conservative. Jump points of $\widehat{G}_y(t)$ can also be randomized, e.g., replace $\widehat{G}_y(t-)$ with $\widehat{G}_y(t-) + U [\widehat{G}_y(t) -\widehat{G}_y(t)]$ in (\ref{eqn:con}) for a $\text{Uniform}(0,1)$ draw $U$, which then  blends the two prediction regions given by using either $\widehat{G}_y(t)$ or $\widehat{G}_y(t-)$ alone.
The conformal prediction method is conservative, but if a randomization scheme is used (whereby the realized prediction interval would depend on the outcome of a random draw,) the method can be made to be exact (cf.~\citealt{vovk2005algorithmic}).

Conformal prediction can be used in the supervised learning setting (i.e., predicting the response given the features as well as labeled training data).
Some recent work includes applying conformal prediction to regression (cf.~\citealt{lei2018distribution}), quantile regression (cf.~\citealt{NEURIPS2019_5103c358}), and lasso regression (cf.~\citealt{lei2019fast}).

\section{Discussion}\label{sec:conclusion}

This paper discusses two major types of methods to construct frequentist prediction intervals.
One is based on an (approximate) pivotal quantity and the other is based on a predictive distribution (or likelihood).
The extensions of these prediction methods to dependent data are briefly discussed.
Here is a summary of our important conclusions.
\begin{itemize}[topsep=0pt,itemsep=-1ex,partopsep=1ex,parsep=1ex]
	\item Exact prediction methods are available for (log-)location-scale distributions under complete and Type II censoring and good approximations are available for Type I censoring.
	\item For (log-)location-scale distributions, there are several equivalent methods for computing {\em exact} intervals. The GPQ predictive distribution method (GPQ-bootstrap) has strong appeal due to its ease of implementation.
	\item For other continuous distributions, the direct-bootstrap method performs no better than the naive plug-in approach and should be avoided (i.e., due to the increased computational costs versus no performance gain over the plug-in method).
	The calibration-bootstrap method, however,
	has good coverage probability properties, even with moderate to small sample sizes.
	Another potentially useful method is to use a (generalized) fiducial predictive distribution.
	\item For discrete distributions, we discussed and illustrated the use of three general methods:
	pivotal cdf (i.e., the conservative method), approximate normal statistics (e.g., based on a Wald-like or a score-like statistic), and integration methods (e.g., a Bayesian method with an objective prior).
	\item When the prediction problems involve dependent data,
	the development of prediction intervals, particularly
	based on parametric bootstrap, requires more investigation for non-Gaussian dependent data.
\end{itemize}

This paper focuses on prediction intervals and coverage probability while in most of the statistical learning (also known as machine learning) literature, the focus is on algorithms for point prediction, which are evaluated with metrics like mean squared error using cross-validation.
However, even with contemporary (nonparametric) prediction algorithms, such as neural networks, boosting, support vector machines, and random forests (cf.~\citealt{clarke_clarke_2018}), there is increasing interest in developing prediction intervals.
In addition to the conformal prediction, for example, prediction intervals based on random forests (cf.~\citealt{zhang2020random}) may be formulated by estimating a distribution of prediction errors (via left out or ``out-of-bag'' observations), similar to the approach described in Section~\ref{subsec:conformal-prediction}.
Ultimately, the development of prediction interval procedures from statistical learning algorithms relates to bridging prediction and estimation, as outlined in the recent overview of \citet{efron2020prediction}.
As our focus in this paper has been on prediction intervals, we have not covered the important area of multivariate prediction.
Recently, some work has been done on multivariate prediction, especially for Gaussian sequence models (cf.~\citealt{george2012minimax} and \citealt{mukherjee2015exact}).

\section*{Acknowledgements}

We would like to thank Professor Jan Hannig for providing a copy of his MATLAB code for computing the Gamma fiducial distribution and Professor Luis A. Escobar for helpful guidance on the use of the pivotal methods for discrete distributions.
We would also like to thank the editor, associate editor, and referees for their constructive and helpful comments.
The research was partially supported by NSF DMS-2015390.

\begingroup
\setlength{\bibsep}{11pt}
\linespread{1}\selectfont
\bibliographystyle{apalike}
\bibliography{reference}  
\endgroup

\end{document}